\documentclass{emulateapj}
\usepackage{psfig,enumerate}
\usepackage{xcolor}
\usepackage{setspace}
\usepackage{color}
\usepackage{atbegshi}
\usepackage{url}
\usepackage[caption=false]{subfig}
\usepackage[pdfpagemode={UseOutlines},bookmarks=true,bookmarksopen=true,
   bookmarksopenlevel=0,bookmarksnumbered=true,hypertexnames=false,
   colorlinks,linkcolor={blue},citecolor={blue},urlcolor={red},
   pdfstartview={FitV},unicode,breaklinks=true]{hyperref}

\begin{document}

\title{APOGEE Chemical Abundance Patterns of the Massive Milky Way Satellites}

\shorttitle{APOGEE Dwarfs}
\shortauthors{Hasselquist et al.}

\author{
Sten Hasselquist\altaffilmark{1,$\dagger$},
Christian R. Hayes\altaffilmark{25},
Jianhui Lian\altaffilmark{1},
David H. Weinberg\altaffilmark{19,37},
Gail Zasowski\altaffilmark{1},
Danny Horta\altaffilmark{23},
Rachael Beaton\altaffilmark{6},
Diane K. Feuillet\altaffilmark{3},
Elisa R. Garro\altaffilmark{7},
Carme Gallart\altaffilmark{34,35},
Verne V. Smith\altaffilmark{24},
Jon A. Holtzman\altaffilmark{18},
Dante Minniti\altaffilmark{7,8}, 
Ivan Lacerna\altaffilmark{12,13},
Matthew Shetrone\altaffilmark{29},
Henrik J\"onsson\altaffilmark{36},
Maria-Rosa L. Cioni\altaffilmark{26},
Sean P. Fillingham\altaffilmark{25},
Katia Cunha\altaffilmark{10,11},
Robert O\'Connell\altaffilmark{2},
Jos\'e G.\ Fern\'andez-Trincado\altaffilmark{12,40},
Ricardo R. Mu\~noz\altaffilmark{39},
Ricardo Schiavon\altaffilmark{23},
Andres Almeida\altaffilmark{16},
Borja Anguiano\altaffilmark{2},
Timothy C. Beers\altaffilmark{9},
Dmitry Bizyaev\altaffilmark{30,31},
Joel R. Brownstein\altaffilmark{1},
Roger E. Cohen\altaffilmark{38},
Peter Frinchaboy\altaffilmark{33},
D. A. Garc{\'i}a-Hern{\'a}ndez\altaffilmark{34,35},
Doug Geisler\altaffilmark{15,16,17},
Richard R. Lane\altaffilmark{20,12},
Steven R. Majewski\altaffilmark{2},
David L. Nidever\altaffilmark{4},
Christian Nitschelm\altaffilmark{22},
Joshua Povick\altaffilmark{4},
Adrian Price-Whelan\altaffilmark{27},
Alexandre Roman-Lopes\altaffilmark{17},
Margarita Rosado\altaffilmark{28},
Jennifer Sobeck\altaffilmark{25},
Guy Stringfellow\altaffilmark{32},
Octavio Valenzuela\altaffilmark{28}
Sandro Villanova\altaffilmark{15},
Fiorenzo Vincenzo\altaffilmark{19}
}


\altaffiltext{1}{Department of Physics \& Astronomy, University of Utah, Salt Lake City, UT, 84112, USA (stenhasselquist@astro.utah.edu)}
\altaffiltext{$\dagger$}{NSF Astronomy and Astrophysics Postdoctoral Fellow}
\altaffiltext{2}{Department of Astronomy, University of Virginia, Charlottesville, VA, 22904, USA}
\altaffiltext{3}{Lund Observatory, Department of Astronomy and Theoretical Physics, Box 43, SE-221\,00 Lund, Sweden}
\altaffiltext{4}{Department of Physics, Montana State University, P.O. Box 173840, Bozeman, MT 59717-3840 (dnidever@montana.edu)}

\altaffiltext{6}{Department of Astrophysical Sciences, Princeton University, 4 Ivy Lane, Princeton, NJ 08544, USA}
\altaffiltext{7}{Departamento de Ciencias Físicas, Facultad de Ciencias Exactas, Universidad Andres Bello, Fernández Concha 700, Las Condes, Santiago, Chile}
\altaffiltext{8}{Vatican Observatory, Vatican City State, V-00120, Italy}
\altaffiltext{9}{Department of Physics and JINA Center for the Evolution of the Elements, University of Notre Dame, Notre Dame, IN  46556, USA}

\altaffiltext{10}{Steward Observatory, The University of Arizona, Tucson, AZ, 85719, USA}
\altaffiltext{11}{Observat\'{o}rio Nacional, 20921-400 So Crist\'{o}vao, Rio de Janeiro, RJ, Brazil}
\altaffiltext{40}{Instituto de Astronom\'ia y Ciencias Planetarias, Universidad de Atacama, Copayapu 485, Copiap\'o, Chile}
\altaffiltext{13}{Millennium Institute of Astrophysics, Nuncio Monsenor Sotero Sanz 100, Of. 104, Providencia, Santiago, Chile}

\altaffiltext{15}{Departamento de Astronom{\'i}a, Casilla 160-C, Universidad de Concepcion, Chile}
\altaffiltext{16}{Instituto de Investigación Multidisciplinario en Ciencia y Tecnología, Universidad de La
Serena. Avenida Raúl Bitrán S/N, La Serena, Chile}
\altaffiltext{17}{Departamento de Física y Astronom{\'i}a, Facultad de Ciencias, Universidad de La Serena. Av. Juan Cisternas 1200, La Serena, Chile}
\altaffiltext{18}{Department of Astronomy, New Mexico State University, Las Cruces, NM 88003, USA}
\altaffiltext{19}{Department of Astronomy, The Ohio State University, 140 W. 18th Ave., Columbus, OH 43210, USA}
\altaffiltext{20}{Centro de Investigaci\'on en Astronom\'ia, Universidad Bernardo
O'Higgins, Avenida Viel 1497, Santiago, Chile}
\altaffiltext{22}{Centro de Astronom{\'i}a (CITEVA), Universidad de Antofagasta, Avenida Angamos 601, Antofagasta 1270300, Chile}
\altaffiltext{23}{Astrophysics Research Institute, Liverpool John Moores University, IC2, Liverpool Science Park, 146 Brownlow Hill, Liverpool L3 5RF, UK}
\altaffiltext{24}{National Optical Astronomy Observatory, 950 North Cherry Ave, Tucson, AZ 85719}
\altaffiltext{25}{Department of Astronomy, University of Washington, Seattle, WA, 98195, USA}
\altaffiltext{26}{Leibniz-Institut f\"{u}r Astrophysik Potsdam, An der Sternwarte 16, D-14482 Potsdam, Germany}
\altaffiltext{27}{Center for Computational Astrophysics, Flatiron Institute, 162 5th Ave., New York, NY 10010, U.S.A.}
\altaffiltext{28}{Instituto de Astronom\'ia, Universidad Nacional Autónoma de M\'exico, A.P. 70-264, 04510, Mexico, D.F., M\'exico}
\altaffiltext{29}{Lick Observatory}
\altaffiltext{30}{Apache Point Observatory and New Mexico State University, Sunspot, NM 88349}
\altaffiltext{31}{Sternberg Astronomical Institute, Moscow State University, Moscow, 119992, Russia}
\altaffiltext{32}{Center for Astrophysics and Space Astronomy, Department of Astrophysical and Planetary Sciences, University of
Colorado, Boulder, CO 80309}
\altaffiltext{33}{Department of Physics \& Astronomy, Texas Christian University, Fort Worth, TX 76129, USA}
\altaffiltext{34}{Instituto de Astrof{\'i}sica de Canarias (IAC), E-38205 La Laguna, Tenerife, Spain}
\altaffiltext{35}{Universidad de La Laguna (ULL), Departamento de Astrof{\'i}sica, E-38206 La Laguna, Tenerife, Spain}
\altaffiltext{36}{Materials Science and Applied Mathematics, Malm\"o University, SE-205 06 Malm\"o, Sweden}
\altaffiltext{37}{Department of Astronomy and Center for Cosmology and AstroParticle Physics, The Ohio State University, 140 W. 18th Ave, Columbus, OH, 43210, USA}
\altaffiltext{38}{Space Telescope Science Institute, 3700 San Martin Drive, Baltimore, MD 21218, USA}

\altaffiltext{39}{Departamento de Astronom\'ia, Universidad de Chile, Camino El Observatorio 1515, Las Condes, Santiago, Chile}

\altaffiltext{12}{Instituto de Astronom\'ia, Universidad Cat\'olica
del Norte, Av. Angamos 0610, Antofagasta, Chile}

\begin{abstract}

The SDSS-IV Apache Point Observatory Galactic Evolution Experiment (APOGEE) survey has obtained high-resolution spectra for thousands of red giant stars distributed among the massive satellite galaxies of the Milky Way (MW): the Large and Small Magellanic Clouds (LMC/SMC), the Sagittarius Dwarf (Sgr), Fornax (Fnx), and the now fully disrupted \emph{Gaia} Sausage/Enceladus (GSE) system. We present and analyze the APOGEE chemical abundance patterns of each galaxy to draw robust conclusions about their star formation histories, by quantifying the relative abundance trends of multiple elements (C, N, O, Mg, Al, Si, Ca, Fe, Ni, and Ce), as well as by fitting chemical evolution models to the [$\alpha$/Fe]-[Fe/H] abundance plane for each galaxy. Results show that the chemical signatures of the starburst in the MCs observed by Nidever et al. in the $\alpha$-element abundances extend to C+N, Al, and Ni, with the major burst in the SMC occurring some 3-4 Gyr before the burst in the LMC. We find that Sgr and Fnx also exhibit chemical abundance patterns suggestive of secondary star formation epochs, but these events were weaker and earlier ($\sim$~5-7 Gyr ago) than those observed in the MCs. There is no chemical evidence of a second starburst in GSE, but this galaxy shows the strongest initial star formation as compared to the other four galaxies. All dwarf galaxies had greater relative contributions of AGB stars to their enrichment than the MW. Comparing and contrasting these chemical patterns highlight the importance of galaxy environment on its chemical evolution.
\end{abstract}


\section{Introduction}

Galaxies are ubiquitous structures in the Universe. While we have made tremendous strides in describing and understanding the patterns seen in galaxies on global or coarsely resolved scales (e.g., mass-metallicity relation, galaxy color-magnitude diagram), our understanding of how these patterns reflect finer details of formation and evolution is severely limited.  This is largely due to the fact that only a few galaxies outside of our own Milky Way (MW) can be studied at the spatial resolution of individual stars, which is required to precisely analyze the galactic star formation histories (SFHs). Fortunately, the Local Group consists of three main massive galaxies (MW, M31, and M33) along with their vast populations of dwarf galaxies, which themselves span a large range in mass, morphology, and environment (e.g., \citealt{Hodge1971,Hodge1989,Mateo1998,Tolstoy2009,McConnachie2012,Ibata2013,Weisz2014,Simon2019}). In principle, detailed SFHs built from photometric or even spectroscopic observations can be constructed for each of these galaxies, allowing for an understanding of the effects that halo mass, formation environment, and interaction history have on a galaxy's SFH.

In practice, the large distance to these galaxies combined with their often large angular size means that fully spatially resolved SFH studies remain observationally costly. Moreover, systematic differences between methods of determining star formation histories can complicate comparisons across multiple galaxies. \citet{Weisz2014} performed a SFH analysis of 40 Local Group dwarf galaxies using photometry from the \emph{Hubble Space Telescope}, deriving SFHs from the color-magnitude diagram (CMD). Although these data did consist of photometry of varying depths (see e.g., \citealt{Ruiz-Lara2018} for effects of photometric depth on SFH determination), the data were all analyzed in a uniform way, and the authors were able to draw reasonably robust conclusions on mass/environmental effects on galaxy evolution. In particular, they found that in comparison to the more massive galaxies, less massive dwarf galaxies generally formed a larger fraction of their stars in the first 2-3 Gyr of their existence. 

The authors also found measurable scatter in SFHs at fixed mass, suggesting where the galaxy formed in relation to other galaxies (i.e., its formation environment) likely has a strong effect on evolution (also see \citealt{Gallart2015}). Many other works find similar SFH scatter at fixed mass (e.g., \citealt{Mateo1998,Grebel1999}), with some galaxies such as Carina experiencing distinct bursts (e.g., \citealt{deBoer2014,Santana2016}). Simulations have proven valuable for understanding exactly how environment affects SFH (e.g., \citealt{Wetzel2016,Revaz&Jablonka2018,Miyoshi2020}), but the extent to which external effects dominate over effects from a galaxy's intrinsic properties is still largely unknown (e.g, \citealt{Kirby2011,Kirby2013,Hendricks2014,Escala2018,Wheeler2019}).

While photometric studies have done much to characterize the star formation rate of Local Group galaxies as a function of time, additional details can be probed with spectroscopic observations from which detailed chemical abundance patterns of individual stars can be obtained. Early star formation efficiencies can be estimated from the ``knee'' in the $\alpha$-element abundance trend (e.g., \citealt{Tinsley1979,Kobayashi2006,Shetrone2003,Nidever2014,Kirby2020}), which has revealed that there is likely a dependence on both galaxy mass and environment on star formation history (e.g., \citealt{Nidever2020}). Additional star formation details, such as variations in the initial mass function (IMF, e.g., \citealt{McWilliam2013,Hasselquist2017,Carlin2018}) or amount of pollution from asymptotic giant branch (AGB) stars (e.g., \citealt{Bonifacio2000,Venn2004,Sbordone2007,Hansen2018,Skuladottir2019,Reichert2020,Fernandez-Trincado2020}) can be probed by close examination of the abundance patterns of hydrostatic/explosive element abundance ratios and $r$-/$s$-process element contributions, respectively. However, the extent to which the abundance patterns can be precisely mapped to parameters that govern star formation (e.g., inflow/outflow, IMF) largely depends on the accuracy of yield tables, which are uncertain for some elements, as well as inherent degeneracies in the predicted model abundance tracks. Moreover, spectroscopic surveys are observationally expensive, and the analysis techniques to extract abundances are susceptible to a range of systematic uncertainties (e.g., using 3D and/or NLTE atmospheres vs. 1D plane-parallel ones). Historically, this has meant that comparative spectroscopic SFH studies have required using heterogeneous data from multiple literature sources.  

Fortunately, the SDSS-III/IV Apache Point Observatory Galactic Evolution Experiment (APOGEE, \citealt{Majewski2017}) has obtained spectra of stars beyond just the Milky Way, including the five most massive MW satellites: Large Magellanic Cloud (LMC), Small Magellanic Cloud (SMC), Sagittarius Dwarf Galaxy (Sgr), Fornax (Fnx), and the \emph{Gaia} Sausage/Enceladus (GSE). The latter is no longer a coherent structure separated from the Milky Way, but its distinct remnant orbit structure means its stars can be relatively easily selected out from the MW (e.g., \citealt{Belokurov2018,Deason2018,Helmi2018,Myeong2018,Gallart2019,Mackereth2019a,Feuillet2020,Horta2021c}), thus serving as a distinct dwarf galaxy for the purposes of this work. These galaxies span two orders of magnitude in mass, and represent a wide range of formation environments, with GSE having merged early with the MW (e.g., \citealt{Gallart2019,Mackereth2019a}), Sgr in the process of merging (e.g., \citealt{Ibata2001,Majewski2003,Belokurov2006,Ruiz-Lara2020}), Fnx in relative isolation now but with some signatures of major mergers in recent times (e.g., \citealt{Amorisco&Evans2012b,delPino2015,delPino2017}), and the MCs falling into the MW for the first time (e.g., \citealt{Besla2007,Kallivayalil2013}), while clearly interacting with each other (e.g., \citealt{Harris&Zaritsky2009,Nidever2010,Besla2016}). Using APOGEE observations, we can now perform a detailed, homogeneous comparison of the SFHs of these galaxies from the \emph{spectroscopic} perspective.

In this work we present the detailed chemical abundance patterns of 10 elements for each of these five galaxies. We quantify the relative differences in the median abundance patterns of these galaxies, allowing for more robust conclusions about their relative star formation efficiencies and SFHs, and more generally about the nucleosynthesis of different elements in these disparate galaxies. We then use two chemical evolution models to interpret these differences as actual physical differences in SFH parameters. Our observations and data reduction are described in \S \ref{sec:obs}, and the sample selection is described in \S \ref{sec:samp}. Chemical Abundance results are shown and compared in \S \ref{sec:results}. We fit chemical evolution models to the abundance results in \S \ref{sec:mod}, and these results are discussed and compared to previous star formation history studies in \S \ref{sec:disc}. Our conclusions are presented in \S \ref{sec:conc}.

\section{Observations and Data Reduction/Analysis}
\label{sec:obs}

Observations were taken as part of APOGEE \citep{Majewski2017}, part of the third and fourth iteration of the Sloan Digital Sky Survey (SDSS-III and SDSS-IV; \citealt{Eisenstein2011} and \citealt{Blanton2017}, respectively). The APOGEE instruments are high-resolution (R $\sim$ 22,000), near-infrared ($H$-band) spectrographs \citep{Wilson2019} observing from both the Northern Hemisphere at Apache Point Observatory (APO) using the SDSS 2.5m telescope \citep{Gunn2006}, and the Southern Hemisphere at Las Campanas Observatory (LCO) using the  2.5m du Pont telescope \citep{Bowen73}. To date, the dual APOGEE instruments have observed some 700,000 stars across the MW and nearby systems, targeting these stars with selections described in \citet{Zasowski2013} and \citet{Zasowski2017}, with updates to the targeting plan described in \citet{Santana2021} and \citet{Beaton2021}. 

We use APOGEE results from the 17th Data Release of SDSS (DR17, Masters et al. in prep). Spectra are reduced as described in \citet{Nidever2015} (with updates described in Holtzman et al. in prep) and analyzed using the APOGEE Stellar Parameters and Chemical Abundance Pipeline (ASPCAP, \citealt{Garcia-Perez2016}), which uses the FERRE code \citep{AllendePrieto2006} to interpolate in a grid of model synthetic spectra \citep{Zamora2015} to find the best-fit stellar parameters and abundances. Updates to the DR17 chemical abundance analysis include cerium abundances using lines characterized in \citet{Cunha2017}, as well as NLTE corrections for Na, Mg, K, and Ca \citep{Osorio2020}. Validation of the APOGEE abundance results can be found in \citet{Jonsson2018}, \citet{Nidever2020}, and Holtzman et al. in prep.

\section{Sample Selection}
\label{sec:samp}

In this work we analyze five dwarf galaxy stellar samples (LMC, SMC, GSE, Sgr, and Fnx) along with a comparison Milky Way sample, for a total of six stellar samples. For each sample, we start with the following cuts:

\begin{itemize}
    \item Median S/N per pixel $>$ 70 (except for Fornax, see \S \ref{app:targ_fnx}) to ensure precise abundances.
    \item No STAR\_BAD bit set in the ASPCAPFLAG\footnote{See DR17 web documentation.} to remove problematic/suspect abundance determinations.
    \item {[Fe/H]} $<$ 0.0 to remove obvious MW contamination from the dwarf galaxy samples.
    \item Remove duplicate observations of a single target that have bit 4 of EXTRATARG set\footnote{See DR17 web documentation for a full description of this flag. https://www.sdss.org/dr16/algorithms/bitmasks/}.
\end{itemize}

Sample selection is then made through a mix of spatial, kinematic, and simple chemical selection criteria. We briefly describe the selection for each sample below, but refer the reader to Appendix A for more detailed plots and descriptions.

\begin{figure*}[t]
\includegraphics[width=1.0\hsize,angle=0]{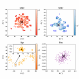}
\caption{Right ascension ($\alpha$) and declination ($\delta$) maps of four galaxies: LMC (red, upper-left), SMC (blue, upper-right), Sgr (orange, lower-left), Fnx (purple, lower-right).  Spatial density bins are plotted except where the bin count is below 5 stars. Black crosses mark the photometric centers of these galaxies, and are noted in Table \ref{tab:select}. The dashed circle in the lower-left panel highlights the Sgr ``main body'' region, the coverage of which is shown in the inset.  About 2/3 of our Sgr targets reside in this main body region. }
\label{fig:map}
\end{figure*}

The criteria used to select each sample are summarized in Table \ref{tab:select}, and the APOGEE IDs for each sample are in Table \ref{tab:members}. To reproduce the results of this paper, one can match Table \ref{tab:members} to the APOGEE allStar catalog\footnote{https://data.sdss.org/datamodel/files/APOGEE\_ASPCAP/APRED\_VERS/ASPCAP\_VERS/allStar.html}. Figure \ref{fig:map} shows the spatial distribution of each galaxy. The APOGEE observations cover much of the spatial extent of these galaxies, including nearly continuous coverage of the Sgr core and stream.

\begin{deluxetable*}{l c c c c c c c r}
\tabletypesize{\scriptsize}
\tablewidth{0pt}
\tablecolumns{3}
\tablecaption{Galaxy Selection \label{tab:select}}
\tablehead{\colhead{System} &  \colhead{Center ($\alpha$, $\delta$)} & \colhead{D$_{\rm proj}$} & \colhead{$V_{\rm helio}$ (km/s)} & \colhead{$\mu_{\alpha}$ (mas/yr)} & \colhead{$\mu_{\delta}$ (mas/yr)}  & \colhead{Photometry} & N$_{*}$ & \colhead{Section}\\}
\startdata
LMC & (80.894$^{\circ}$,-69.756$^{\circ}$) & $<$ 12$^{\circ}$ & 161 $<$ $V_{\rm helio}$ $<$ 370 & 1.01 $<$ $\mu_{\alpha}$ $<$ 2.62 & -1.15 $<$ $\mu_{\delta}$ $<$ 1.70 & (J-K$_{s}$) $<$ 1.3; RGB\footnote{RGB tip, described in \S \ref{app:targ_mc}} & 3,909 & \ref{sec:targ_mc};\ref{app:targ_mc}\\
SMC & (13.187$^{\circ}$,-72.829$^{\circ}$) & $<$ 8$^{\circ}$ & 66 $<$ $V_{\rm helio}$ $<$ 235 & 0.05 $<$ $\mu_{\alpha}$ $<$ 1.51 & -1.57 $<$ $\mu_{\delta}$ $<$ -0.94 & (J-K$_{s}$) $<$ 1.3; RGB & 1,146 & \ref{sec:targ_mc};\ref{app:targ_mc}\\
GSE\footnote{Primarily orbital selections} & $-$ & $-$ & $-$ & $-$ & $-$ & $-$ & 972 & \ref{sec:targ_gse};\ref{app:targ_gse}\\
Sgr\footnote{Primarily kinematic selections in Sgr coordinate frame} & $-$ & $-$ & $-$ & $-$ & $-$ & $-$ & 946 &\ref{sec:targ_sgr};\ref{app:targ_sgr}\\
Fnx & (39.748$^{\circ}$, -34.376$^{\circ}$) & $< 0.9^{\circ}$\footnote{Single APOGEE plug plate} & 17 $<$ $V_{\rm helio}$ $<$ 89 & 0.17 $<$ $\mu_{\alpha}$ $<$ 0.60 & -0.71 $<$ $\mu_{\delta}$ $<$ -0.05 &  $-$ & 192 & \ref{sec:targ_fnx};\ref{app:targ_fnx}\\
\enddata
\end{deluxetable*}

\begin{deluxetable}{l r}
\tabletypesize{\scriptsize}
\tablewidth{0pt}
\tablecolumns{3}
\tablecaption{Galaxy Selection \label{tab:members}}
\tablehead{\colhead{APOGEE ID} & \colhead{System} \\}
\startdata
2M03141881-7642442 & LMC \\
2M03173277-7702254 & LMC \\
2M03174293-7712511 & LMC \\
...\footnote{full version online} 
\enddata
\end{deluxetable}

\subsection{Magellanic Clouds}
\label{sec:targ_mc}

The Magellanic Clouds have been extensively observed by APOGEE-2S, with many programs targeting stars across a wide range of evolutionary type (see \citealt{Nidever2020} and \citealt{Santana2021}), including red giant branch (RGB) stars, oxygen-rich asymptotic giant branch (O-AGB) stars, C-rich AGB stars (C-AGB), and massive (M $\gtrsim$ 3 M$_{\odot}$) red supergiant (RSG) stars. To select our MC sample we use a combination of spatial, kinematical, color, magnitude, and metallicity cuts, described in detail in \S \ref{app:targ_mc}. Because only the RGB stars have well-vetted stellar parameters and abundances from APOGEE, we remove the O-AGB, C-AGB, and RSG stars by only selecting stars that are below the Red Giant Branch (RGB) tip (as defined by \citealt{Hoyt2018}).  However, because the stars above the RGB tip are generally the more massive stars in the MCs, removing these stars means that we are biasing our sample against the youngest (age $\lesssim$ 1 Gyr) MC stars. These cuts result in samples of $\sim$3,900 stars for the LMC and $\sim$1,100 stars for the SMC. This sample is largely comprised of RGB stars, but likely still contains AGB stars that are below the RGB tip. These stars span a large spatial extent of the MCs, as shown in the top row of Figure \ref{fig:map}. 

The log(g)-T$_{\rm eff}$ distribution of these stars are shown in the left two panels of Figure \ref{fig:hr}. While most of the MC stars that pass these cuts have log(g) and T$_{\rm eff}$ consistent with being upper giant branch stars, the LMC contains three groups of stars that have different stellar parameters than the majority of the sample: (1) the clump of log(g) $<$ 0 stars in the LMC panel of Figure \ref{fig:hr}, (2) the cool, higher gravity stars (log(g) $>$ 1.5 \& T$_{\rm eff}$ $<$ 4200 K, and (3) stars at T$_{\rm eff}$ $>$ 5000 K. After analyzing these stars and comparing to other surveys, we have determined that (1) are thermally pulsating asymptotic giant branch stars (TP-AGB), enhanced in C+N, (2) are likely mass-transfer binaries, enhanced in [Ce/Fe], and (3) are identified as Delta Cep pulsators by \citet{soszynski2008}. We do not explicitly remove these stars, and present their chemical results in \S \ref{sec:chem_results}.

\subsection{Gaia Sausage/Enceladus (GSE)}
\label{sec:targ_gse}

In this work, we treat the accreted halo stars, defined below, as originating from one progenitor, the GSE, that has since merged with the MW. However, when treating the GSE as one entity, we assume that we have selected stars in a way that is not chemically biased. A compact remnant for GSE has not yet been confirmed. Should this remnant exist and be absent from our selection, then it is possible we are missing the most chemically evolved GSE stars in our sample. Furthermore, like other studies, we are assuming the stars come from one progenitor rather than multiple. We are also only explicitly removing stars in known globular clusters, but should GSE contain dissolved GC stars, then we might expect our sample to contain stars that show GC-like abundance pattern variations. However, these stars are likely to comprise only a small fraction of our GSE sample, and will not significantly impact our interpretations of the median abundance trends.

\begin{figure*}[t]
\includegraphics[width=1.0\hsize,angle=0]{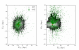}
\caption{Map of the GSE members (green) studied in this work using astroNN distances \citep{Leung2019}. Left: Rectangular X and Y Galactic coordinates, where the position of the Sun is notated by a red cross. Right: vertical height above/below the plane plotted as a function of Galactic cylindrical radius. The grey-scale density map shows the location of APOGEE MW stars.}
\label{fig:map_gse}
\end{figure*}

There are now a wide range of literature works that show how GSE stars can be selected by applying various kinematical and dynamical selection criteria. In this work, we follow \citet{Feuillet2020} and select stars in $\sqrt{J_{R}}$-L$_{z}$ space, with these quantities provided in the astroNN\footnote{https://www.sdss.org/dr16/data\_access/value-added-catalogs/?vac\_id=the-astronn-catalog-of-abundances,-distances,-and-ages-for-apogee-dr16-stars} APOGEE DR17 value-added catalog\citep{Leung2019}. These selections result in some contamination from the high-$\alpha$ MW disk, so we perform a chemical cut in [(C+N)/Fe] for stars with [Fe/H] $>$ -1.05 to remove the obvious high-$\alpha$ MW stars, which are $\sim$ 0.2-0.3 dex enhanced in [(C+N)/Fe] as compared to GSE. Selecting GSE members is described in more detail in \S \ref{app:targ_gse}, and our final sample consists of $\sim$ 1000 stars. Figure \ref{fig:map_gse} shows that our GSE sample studied here primarily comes from stars near the solar radius, at $\pm$5 kpc from the plane of the MW, although some stars do come from much further away. To make these maps we use distances from DR17 astroNN \citep{Leung2019}. 

\subsection{Sagittarius}
\label{sec:targ_sgr}

To select Sgr members, we take a similar approach to that of \citet{Hayes2020}. We first only consider stars with [Fe/H] $<$ 0.0, heliocentric distance greater than 10 kpc, and within $\pm$30$^{\circ}$ of the plane of the Sgr stream \citep{Majewski2003}. Then, we make an initial selection in $V_{\rm zs}$ - $L_{\rm zs}$ plane, where $V_{\rm zs}$ and $L_{\rm zs}$ are the vertical velocity and angular momenta in the Sagittarius Galactocentric coordinate system, as derived and described in \citet{Majewski2003}. We then make further selections in the $\phi_{\rm vel,s}$-$\Lambda_{\rm s}$ plane, where $\phi_{\rm vel,s}$ is the velocity direction in the X and Y directions of the Sgr coordinate system and $\Lambda_{\rm s}$ is the longitude along the Sgr stream (\citealt{Majewski2003,Hayes2020}), to remove stars that are moving perpendicular to the stream. We describe the selection process in more detail in \S \ref{app:targ_sgr}. The final Sgr sample consists of $\sim$1,000 stars, $\sim$ 2/3 of which are in the main body of Sgr (within 12$^{\circ}$ of the center of Sgr). See $\S$ \ref{app:targ_sgr} for details on the coverage of the Sgr main body.

\subsection{Fornax}
\label{sec:targ_fnx}

The APOGEE Fornax field was designed specifically to observe likely members based on radial velocities, proper motions, and/or CMD (see \citealt{Zasowski2017} and \citealt{Santana2021}). Therefore, most targets in this field are likely Fornax members, but we re-analyze the APOGEE RVs and \emph{Gaia} proper motions to remove any contamination.  We adopt a lower cut of S/N $>$ 40 for the Fornax sample to include a meaningful number of stars to compare to the other galaxies. This lower S/N cut means that the individual abundance uncertainties are generally larger for Fnx stars than the stars in other galaxies. The final selection cuts we used are shown in Table \ref{tab:select}, and the process is described in more detail in \S \ref{app:targ_fnx}.

\subsection{MW Comparison Sample}

\begin{figure*}[t]
\includegraphics[width=1.0\hsize,angle=0]{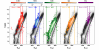}
\caption{HR Diagram for the five different galaxy samples (monochromatic density maps and circles where bins contain fewer than 5 stars) plotted with the HR diagram of the T$_{\rm eff}$-matched MW stars (grayscale density). Horizontal lines indicate the 2$\sigma$ T$_{\rm eff}$ range adopted to select out the MW comparison samples. }
\label{fig:hr}
\end{figure*}

While the APOGEE pipeline has had several improvements to eliminate T$_{\rm eff}$ and log(g) systematics in abundance determination \citep{Jonsson2020}, recent work by \citet{Griffith2020} shows that some elemental abundances still exhibit some small systematic trends in abundance with T$_{\rm eff}$ and log(g) (e.g., Al and Si). Thus, to minimize these effects on our interpretations of the abundance trends, we compare each galaxy to a MW sample of roughly similar stellar parameters. For each galaxy, we select a T$_{\rm eff}$ range corresponding to $\pm 2\sigma$ from the median T$_{\rm eff}$ values of each galaxy, as shown in Figure \ref{fig:hr}.

For the MCs, Sgr, and Fnx, we are primarily analyzing the abundance patterns of luminous giants (log(g) $\lesssim$ 1.5). While we do not make specific spatial selections for the MW stars, the MW comparison sample for the MCs, Sgr, and Fnx, covers much of MW disk and bulge region. The GSE sample spans much of the giant branch (0.5 $<$ log(g) $<$ 3.0), so the MW comparison sample contains a larger fraction of intrinsically less-luminous stars, resulting in a MW sample that is primarily located spatially within 1-2 kpc from the Sun.  \citet{Weinberg2019} show that the median trends of APOGEE elemental abundance ratios are nearly independent of location in the disk provided one separates the low-$\alpha$ and high-$\alpha$ populations, so we do not expect geometrical selection effects within the MW to have an effect on our comparison.

\section{Chemical Abundance Results}
\label{sec:results}

In this section, we present the elemental abundances for C+N, the $\alpha$-elements (O, Mg, Si, Ca), Al, Ni, and Ce. We select these elements because they are among the most precise APOGEE abundances across the full parameter space covered here, and are among the most accurate when comparing to optical studies (see e.g., \citealt{Jonsson2018,Jonsson2020}). We combine C and N because stars will change their [C/Fe] and [N/Fe] abundances during dredge up ascending to and along the giant branch, but these processes occur in such a way that the [(C+N)/Fe] abundance is largely constant before and after these mixing processes (see e.g., \citealt{Gratton2000}). We compare the abundance patterns of each galaxy to the abundance pattern of the MW in this section, as well as compare the abundance patterns of the dwarf galaxies to each other.

\begin{figure*}[t]
\includegraphics[width=0.9\hsize,angle=0]{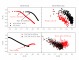}
\caption{Chemical evolution model tracks of the [Si/Fe]-[Fe/H] abundance plane (left column) along with mock observations sampled from these tracks with 0.05 dex abundance uncertainties (right column). The top row compares two galaxy model tracks with different early star formation efficiency, and the bottom row compares two galaxy model tracks with identical early star formation efficiencies, but one with a secondary burst of star formation.}
\label{fig:schem}
\end{figure*}

Throughout these sections, we describe various aspects of the abundance patterns of each galaxy, linking certain features to physical drivers of SFH, such as early star formation efficiency or presence/strength of a secondary star burst. We provide an example schematic diagram of how we interpret abundance patterns in Figure \ref{fig:schem}. In the top row of Figure \ref{fig:schem} we show flexCE (described in detail in \citealt{Andrews2017}, \S \ref{sec:flexce}, and \S \ref{app:modeling}) chemical evolution model tracks (left) with mock observations of the tracks (right) to show what the abundance patterns of two galaxies (labeled as ``Dwarf Galaxy Model'' and ``MW Model'' in the Figure) with different initial star formation efficiencies look like in the [Si/Fe]-[Fe/H] abundance space. The combination of the star formation efficiency (SFE $\equiv$ $\dot{\rm M}_{*}/{\rm M}_{\rm gas}$) and the gas supply (${\rm M}_{\rm gas}(t)$) determines the star formation rate (SFR$(t) = \dot{\rm M}_{*}(t)$). The gas supply in turn depends on the gas accretion history, and is further regulated by star formation and outflows (which deplete ${\rm M}_{\rm gas}$) and recycling from evolved stars (which replenishes ${\rm M}_{\rm gas}$).

The top row of Figure \ref{fig:schem} shows that the [Si/Fe]-[Fe/H] abundance pattern of the lower SFE dwarf galaxy model (red) begins to decrease with increasing [Fe/H] at much lower [Fe/H] values than the higher SFE MW model (black). In both galaxies, the [Si/Fe] abundance decreases after Type Ia SNe begin to substantially contribute to the chemical evolution of these galaxies, which is thought to occur some time after star formation begins (see e.g., \citealt{Tinsley1979}). Type Ia SNe produce Fe without producing much Si (which is primarily produced in Type II SNe), resulting in a decrease of the [Si/Fe] abundance as the Type Ia/Type II SNe ratio increases. Because the delay time at which Type Ia SNe begin to contribute to the chemical enrichment of these galaxies is assumed to be the same, the metallicity at which [Si/Fe] begins to decrease (sometimes referred to as the $\alpha$-element abundance ``knee'') probes the early SFE of a galaxy, with more metal-poor knees indicating a galaxy experienced fewer Type II SNe events contributing to its enrichment before the delayed Ia SNe started to contribute. Therefore, from these abundance patterns, we would conclude that the dwarf galaxy experienced lower efficiency SF at early times than the MW. 

The bottom row of Figure \ref{fig:schem} compares two dwarf galaxies with the same early star formation efficiency, but the red track emphasizes how a starburst influences the [Si/Fe]-[Fe/H] track. The red model track was used in \citet{Nidever2020} to explain the rising [$\alpha$/Fe] abundance pattern observed in the APOGEE LMC data. We would therefore conclude that the red dwarf galaxy experienced some secondary star formation epoch whereas the black dwarf galaxy did not. In this example, the starburst is induced by temporarily increasing the star formation efficiency while maintaining the same gas accretion history (see \S \ref{sec:flexce}). Many more examples of the influence of star formation and outflow parameters on evolutionary tracks can be found in \citet{Andrews2017} and \citet{Weinberg2017}, and a systematic exploration of chemical evolution tracks with starbursts can be found in \citet{Johnson2020}.

\begin{figure*}[t]
\includegraphics[width=0.9\hsize,angle=0]{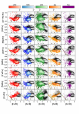}
\caption{Chemical abundance patterns of select elements for each of the dwarf galaxies as compared to the MW. Over-plotted for each galaxy are 2D density histograms except for bins where the density falls below 5 stars. The dashed black line in the [Al/Fe], [Ni/Fe], and [Ce/Fe] panels indicate the grid edges, below which the APOGEE spectra grids do not extend. Representative median individual uncertainties in the abundances for each dwarf galaxy are shown in black in the bottom of each panel.}
\label{fig:mw_comp}
\end{figure*}

\begin{figure*}[t]
\includegraphics[width=0.9\hsize,angle=0]{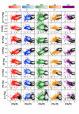}
\caption{Same as in Figure \ref{fig:mw_comp}, but using Mg as the reference element instead of Fe.}
\label{fig:mw_comp_mg}
\end{figure*}

APOGEE abundance results of each dwarf galaxy as compared to their respective MW comparison samples are shown in Figures \ref{fig:mw_comp} and \ref{fig:mw_comp_mg}. Figure \ref{fig:mw_comp} shows the ``traditional'' abundance patterns using Fe as the reference element ([X/Fe]-[Fe/H]), whereas Figure \ref{fig:mw_comp_mg} shows the abundance patterns using Mg as a reference element ([X/Mg]-[Mg/H]) to analyze abundance patterns as a function of Type II SNe ejecta alone, following \citet{Weinberg2019}. Each row of these Figures shows a different elemental abundance ratio ([X/Fe] or [X/Mg]) plotted against a ``metallicity indicator'', represented by [Fe/H] or [Mg/H]. Each column shows how a given dwarf galaxy's abundances compare to that of its respective MW comparison sample. The MW comparison samples are nearly identical for the LMC, SMC, Sgr and Fnx panels, but the MW comparison sample for GSE contains a much larger fraction of lower luminosity stars. The chemical abundance pattern for each sample is plotted as a density map except for where the pixel contains fewer than 5 stars, where the individual measurements are displayed as circles instead.

\begin{figure*}[t]
\includegraphics[width=1.0\hsize,angle=0]{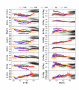}
\caption{Median abundance tracks (solid colored lines) and associated $\pm$1 $\sigma$ uncertainties on the median track (colored shaded regions) for each galaxy in moving bins of 30 stars. The MW sample is plotted as a grayscale density image. We have removed all GSE and MW stars with $T_{\rm eff} >$ 4200 K to make for relatively systematic-free comparison across all galaxies. Note that vertical axis ranges vary from panel to panel, but grey dot-dashed horizontal lines are spaced at 0.20 dex in all panels so they can be used as a visual reference for comparing the strength of metallicity trends across elements.}
\label{fig:medians}
\end{figure*}

To compare the chemical tracks of the dwarf galaxies with each other, we also show the median abundance tracks for the dwarf galaxies all in one panel per chemical element in Figure \ref{fig:medians}, with [X/Fe]-[Fe/H] plotted on the left column and [X/Mg]-[Mg/H] plotted on the right column. Running medians are calculated in bins of 30 stars. Because GSE sampled a much larger section of the giant branch than the other dwarf galaxies, we only include stars with 3600 K $<$ T$_{\rm eff} < 4200$ K to make for a more systematic-free comparison. We also select a comparison MW sample to now only cover 3600 K $<$ T$_{\rm eff} < 4200$ K.

In the following subsections, we describe the chemical abundance patterns of each galaxy, regularly referring to Figures \ref{fig:mw_comp}, \ref{fig:mw_comp_mg}, and \ref{fig:medians}.

\subsection{LMC}
\label{sec:results_lmc}

The left columns of Figures \ref{fig:mw_comp} and \ref{fig:mw_comp_mg} show the abundance patterns for the LMC. The APOGEE LMC sample contains stars across a wide range of metallicities, -2.2 $<$ [Fe/H] $<$ -0.3, making it currently one of the most metal-rich MW satellite galaxies (compare the median abundance tracks in Figure \ref{fig:medians}). Such a wide metallicity range implies an extended star formation history, the complexity of which is highlighted in the [X/Fe] and [X/Mg] abundance patterns, as described in detail below. 

\subsubsection{O, Mg, Si, and Ca}

The $\alpha$-elements, O, Mg, Si, and Ca, are primarily produced in massive stars and released to the ISM via Type II SNe, with Si and Ca having non-negligible contributions from Type Ia SNe in the MW disk (e.g., \citealt{Weinberg2019}). As already shown by \citet{Nidever2020}, and again presented in the left columns of Figures \ref{fig:mw_comp} and \ref{fig:medians}, the APOGEE LMC sample exhibits an [$\alpha$/Fe]-[Fe/H] abundance pattern that decreases over the metallicity range -2.2 $<$ [Fe/H] $<$ -1.2 to sub-solar [$\alpha$/Fe] values, before increasing to intersect with the MW low-$\alpha$ disk trend. \citet{Nidever2020} interpreted this pattern as very weak (i.e., low efficiency) early star formation, followed by a strong burst in star formation that occurred in more recent times, when many Type II SNe drove up the $\alpha$-element abundance. This picture is qualitatively consistent with photometric studies of the LMC (e.g., \citealt{Harris&Zaritsky2009,Monteagudo2018,Ruiz-Lara2020b}), but is only somewhat consistent with other spectroscopic studies. Many spectroscopic abundance studies (e.g., \citealt{Smith2002,Lapenna2012,VanderSwaelmen2013}) find that the more metal-poor LMC stars are deficient in the $\alpha$ elements relative to the MW, but only the Mg abundances of \citet{Lapenna2012} show a flat or slightly increasing [Mg/Fe] abundance with metallicity at [Fe/H] $>$ -1.5. We refer the reader to \citet{Nidever2020} for a more detailed discussion of the comparison samples, but note here that the APOGEE sample studied in this work is a factor of $\sim$40 larger than other high-resolution abundance studies. Moreover, the APOGEE random uncertainties of individual abundance measurements are $\sim$1/4 those presented in other studies.

While the pattern of low early SF followed by a major burst is seen across all $\alpha$-elements, the extent to which the metal-rich LMC stars overlap the MW disk sample varies. The [O/Fe] and [Mg/Fe] abundance patterns reach the MW low-$\alpha$ disk trend, whereas the [Si/Fe] and [Ca/Fe] actually rise above the MW low-$\alpha$ disk trend (perhaps best seen in the left column of Figure \ref{fig:medians}). The [O/Mg], [Si/Mg], [Ca/Mg] abundances all slightly decrease with increasing [Mg/H], with [Ca/Mg] remaining $\sim$0.05 dex above the MW trend at [Mg/H] = -0.4.

Compared to the other dwarf galaxies (Figure \ref{fig:medians}), only Sgr extends to as high metallicities as the LMC. However, the metal-poor LMC stars have lower [$\alpha$/Fe] (by $\sim$ 0.05-0.1 dex) than the Sgr stars until [Fe/H] = -1.0, at which point the LMC increases its [$\alpha$/Fe] abundance, whereas the [$\alpha$/Fe]-[Fe/H] abundance track of Sgr continues to decrease before becoming mostly flat. Over the metallicity range -1.5 $<$ [Fe/H] $<$ -1.0, the LMC is $\sim$ 0.2 dex deficient across all $\alpha$ elements compared to GSE. Despite these differences in the [$\alpha$/Fe]-[Fe/H] tracks, most galaxies show similar decreasing [$\alpha$/Mg] abundances with increasing [Mg/H], the exception being Fornax in [Ca/Mg] (discussed more in \S \ref{sec:results_fnx}).

\subsubsection{Carbon and Nitrogen}

Carbon and nitrogen are elements that are thought to be produced in great quantities in type II SNe, with nitrogen yields expected to have some dependence on progenitor metallicity (e.g., \citealt{Kobayashi2006}). Both elements are also thought to be produced in appreciable amounts by AGB stars (e.g., \citealt{Nomoto2013,Karakas2014,Andrews2017,Rybizki2017}). Throughout this work, we analyze the sum of these abundances, [(C+N)/Fe] and [(C+N)/Mg], as red giant stars undergo dredge up processes that mix the surface abundances with material produced as a consequence of nuclear reactions deeper in the star. This mixing operates in a way such that the star's birth C+N abundance is largely conserved (e.g., \citealt{Iben1964,Gratton2000,Salaris2015,Vincenzo2021}). 

The [(C+N)/Fe] abundance pattern of the LMC shows a slight decrease from -0.1 to $\sim$ -0.3 dex over the range -2.2 $<$ [Fe/H] $<$ -1.2. At [Fe/H] $>$ -1.2, the [(C+N)/Fe] is flat at ~0.3 dex below the MW abundance trend before rapidly increasing to almost intersect with the MW low-$\alpha$ ``thin'' disk trend at [Fe/H] = -0.3. In the [(C+N)/Mg]-[Mg/H] abundance plane shown in Figure \ref{fig:mw_comp_mg} and in the right column of Figure \ref{fig:medians}, both the LMC and the MW high-$\alpha$ ``thick'' disk stars show similar trends of slightly increasing [(C+N)/Mg] with [Mg/H] in their region of overlap, with the LMC pattern being enhanced by $\sim$ 0.1 dex. Over the wide metallicity range (-1.7 $<$ [Mg/H] $<$ -0.5) the LMC [(C+N)/Mg] trend is flat to within $\sim$ 0.05 dex, implying that (C+N) production tracks Mg production at these metallicities. 

As shown in Figure \ref{fig:medians}, only Sgr appears to show a similar increase in [(C+N)/Fe] as the LMC at [Fe/H] $>$ -0.7, although this increase is not as steep as the increase in the LMC. Most other galaxies show a similar slight decrease of [(C+N)/Fe] from [Fe/H] = -2.0 to [Fe/H] = -1.0, although both Fnx and GSE have enhanced [(C+N)/Fe] over this metallicity range as compared to the LMC by $\sim$0.15 dex. However, the [(C+N)/Mg]-[Mg/H] abundance patterns show that Fnx is enhanced as compared to the LMC, and GSE is actually slightly deficient.

\subsubsection{Aluminum and Nickel}

Aluminum is an element thought to be produced by massive stars and released to the ISM solely via Type II SNe with some dependence on progenitor metallicity (e.g., \citealt{Weinberg2019}), whereas Nickel is an element produced in both Type II and Type Ia SNe. The [Al/Fe]-[Fe/H] abundance pattern of the LMC appears qualitatively consistent with the starburst scenario (flat, then increasing abundance at point of star burst). The [Al/Mg] plot shown in the left column of Figure \ref{fig:mw_comp_mg} shows that the [Al/Mg] ratio is still about 0.2 dex below the MW trend at the same [Mg/H]. The [Al/Mg]-[Mg/H] abundance tracks are similar across all of the dwarf galaxies (right column of Figure \ref{fig:medians}), showing a steady rise from [Al/Mg] $\simeq$ -0.6 at [Mg/H] $\simeq$ -1.7 to [Al/Mg] $\simeq$ -0.25 at [Mg/H] $\simeq$, which suggests that Type II Al yields increase steadily over this metallicity range. 

The LMC Ni abundance trends, for both [Ni/Fe] and [Ni/Mg], are similar to the $\alpha$-element abundance trends, being most similar to [Si/Fe] and [Si/Mg].  However, [Ni/Fe] remains below the MW low-$\alpha$ disk trend whereas the $\alpha$ elements reach the same level as the MW low-$\alpha$ disk trend at the highest LMC metallicities. The decline of [Ni/Mg] with increasing [Mg/H] is also steeper than that of [Si/Mg], probably because of a larger Type Ia SNe contribution. The [Ni/Mg] of the metal-rich LMC is slightly deficient as compared to the metal-rich Sgr stars, with the LMC stars lying in between the two MW sequences, and Sgr lying closer to the MW low-$\alpha$ sequence.

\subsubsection{Cerium}

The heavy $s$-process element cerium is one of the elements whose abundances
are presented in DR17 and is based upon Ce II lines as described in \citet{Cunha2017}.  The astrophysical source of the s-process elements, such as
Ce, is dominated by thermally-pulsing (TP) AGB stars, with the probable
neutron source being $^{13}$C($\alpha$,n)$^{16}$O (e.g., \citealt{Karakas2016,Prantzos2018,Kobayashi2020}). This particular neutron source is important, as it takes place in the inter-shell
region (between shell He-burning triple-$\alpha$ and shell H-burning
via CN-cycle reactions) between thermal pulses (e.g., \citealt{Karakas2014}) 
and is a primary
neutron source (i.e., independent of the star's birth metallicity), driven by the mixing of protons into $^{12}$C-rich regions
(from triple-alpha) resulting in $^{12}$C(p,$\gamma$)$^{13}$N($\beta^{+}$,$\nu$)$^{13}$C.
Cerium is a sensitive diagnostic of TP-AGB chemical enrichment, which continues 
to take place over relatively long timescales (of order several Gyr), and provides 
information on nucleosynthesis and chemical evolution that complements that
provided by the $\alpha$-elements (massive stars and Type II SNe), the Fe-peak
(Type Ia SNe), or the $r$-process (merging neutron stars).  

The LMC [Ce/Fe] pattern increases with [Fe/H] across nearly the entire metallicity range, with the most metal-rich MW stars having slightly enhanced [Ce/Fe] as compared to the MW disk. The rise
of [Ce/Fe] with [Fe/H] is similar to what was found by \citet{Pompeia2008} in
a sample of LMC red giants for the s-process element barium, which will behave
similarly to Ce. There is also a small clump of stars at [Fe/H] $\simeq$ -0.7 that are $\sim$ 0.2 dex more enhanced than the rest of the LMC stars, which pulls the median up in Figure \ref{fig:medians} to be slightly above the [Ce/Fe] abundance pattern of Sgr. 

The LMC [Ce/Mg]-[Mg/H] abundance pattern also shows an increasing [Ce/Mg] with [Mg/H], however, the lower-left panel of Figure \ref{fig:mw_comp_mg} and the lower-right panel of Figure \ref{fig:medians} shows that the [Ce/Mg] increases more slowly with [Mg/H] at [Mg/H] $>$ -1.0. In the regime of $-1.8 <$ [Mg/H] $< -0.7$, the rise of [Ce/H] with [Mg/H] is well-fit by a slope of 1.4,
while in the range of -0.7 $<$ [Mg/H] $<$ -0.2 the slope drops to $\sim$0.9. The different behavior from [Ce/Fe] arises because [Mg/Fe] itself is rising over this interval, probably because of a burst in star formation. This burst evidently boosts Mg more rapidly than Ce, which is as expected if Ce production is dominated by less massive stars. Such a change in slope is less evident in the [Ce/Mg]-[Mg/H] abundance patterns of Sgr, which has a flat [Mg/Fe] trend at high metallicity.

\subsubsection{LMC Interpretations}

The metal-poor $\alpha$-element abundance patterns show that, compared to GSE, Sgr and the MW, the LMC enriched to a much lower [Fe/H] before Type Ia SNe began to contribute to its enrichment, suggesting much lower early star formation efficiency. We interpret the increasing $\alpha$-element abundance patterns at higher metallicity, along with the increases of both [(C+N)/Fe] and [Al/Fe] with increasing [Fe/H], as results of a major starburst in the LMC that occurred at [Fe/H] $\simeq$ -0.8 (see \S \ref{sec:chem_results} for quantitative modeling). The [Ni/Fe]-[Fe/H] abundance pattern of the LMC is qualitatively similar to the $\alpha$-element abundance tracks (slight decrease at the metal-poor end followed by a slight increase from the burst), but [Ni/Fe] in the LMC remains below the MW trend at [Fe/H] $>$ -0.7, whereas the $\alpha$ elements all reach the MW low-$\alpha$ disk trend. Because the [Ni/Mg] ratio at this point in the LMC's evolution is closer to that of the high-$\alpha$ MW disk (e.g., the ``pure'' Type II SNe [Ni/Mg] abundance), the expectation is that the [Ni/Fe] should be closer to the abundance of the high-$\alpha$ MW disk. The [Ni/Fe] deficiency is therefore plausibly a result of the LMC lacking more metal-rich Type Ia SNe contributing its chemical evolution, if the production of Ni in Type Ia SNe is indeed metallicity-dependent (see e.g., \citealt{Woosley&Weaver1995,Chieffi2004,Seitenzahl2013,Andrews2017,Rybizki2017}).

While the trend of increasing [Al/Mg] with increasing [Mg/H] in all dwarf galaxies can be explained by metallicity-dependent Type II SNe yields, the slight deficiency in [Al/Mg] for the most metal-rich LMC and Sgr stars relative to MW stars is a bit difficult to explain given that Al and Mg are both pure type II SNe products (e.g., \citealt{Weinberg2019}). One plausible explanation is that the low star formation efficiencies in the dwarfs cause a stronger ``metallicity lookback'' effect -- e.g., when the metallicity of the ISM (and newly forming stars) is -0.5, the average metallicities of the stars that enriched the ISM is still significantly lower than -0.5, with consequently lower Al production. This could also be an explanation for the Ni abundance patterns if Ni production is metallicity-dependent in both Type Ia and Type II SNe. 

[O/Mg] and [Si/Mg] for the LMC, and other dwarf galaxies, slowly decrease (by about 0.1 dex) over the range -1.3 $<$ [Mg/H] $<$ -0.5, with both the LMC and Sgr joining the MW trends at [Mg/H] $>$ -0.5. This is either a result of metallicity-dependent Type II SNe yields for these elements (e.g., more Si and O relative to the MW at low metallicity), or a result of a slowly increasing Type II/Type Ia SNe fraction, as star formation was extended after the initial SF epoch. The [Fe/Mg] suggests the latter scenario is plausible, at least from -0.8 $<$ [Mg/H] $<$ -0.4, over which the [Fe/Mg] ratio slowly decreases as more Type II SNe contribute. The slightly enhanced [Ca/Mg] for the LMC and Sgr at these metallicities as compared to the MW stars is a result of Type Ia contribution to Ca, which is still substantial even with the starburst injecting many more Type II SNe products than at lower metallicities. These results suggest that Ca has a higher contribution from Type Ia SNe than Si (also see e.g., \citealt{Tsujimoto1995,Hayes2018a}).

We interpret the slightly increasing [(C+N)/Mg] with [Mg/H] as metallicity-dependent C+N yields in Type II SNe (since the MW high-$\alpha$ stars also show this trend), combined with some contribution of C+N from another source, such as AGB stars, which results in the LMC (and Sgr) trend being slightly elevated from the MW high-$\alpha$ sequence. However, this is a small effect, with [(C+N)/Mg] varying by only $\pm$0.05 dex over the metallicity range -1.7 $<$ [Mg/H] $<$ -0.3. The enhanced [Ce/Fe] and [Ce/Mg] both suggest that the LMC and Sgr experienced significant contributions to their evolution from the $s$-process (likely from AGB stars), with the most metal-rich stars enhancing to similar levels of [Ce/Mg] as the more metal-rich MW low-$\alpha$ sequence. The slight change in slope observed in the [Ce/Mg]-[Mg/H] abundance pattern of the LMC is likely due to the starburst polluting the ISM with much more Mg from Type II SNe.

\subsection{SMC}
\label{sec:results_smc}

The second column of Figures \ref{fig:mw_comp} and \ref{fig:mw_comp_mg} show the abundance patterns of the SMC, and the median abundance tracks (blue) are compared to the other galaxies in Figure \ref{fig:medians}. The SMC only gets as metal-rich as [Fe/H] $\simeq$ -0.6 and [Mg/H] $\simeq$ -0.7, $\sim$ 0.3-0.4 dex less than the LMC. This is an expectation from the established mass-metallicity relation in Local Group dwarf galaxies (e.g., \citealt{Kirby2010}), where the more massive galaxies tend to be more metal-rich. However, the most metal-rich stars of the SMC are still $\sim$ 0.5 dex more metal poor than the most metal-rich stars in Sgr.

\subsubsection{O, Mg, Si, and Ca}

The $\alpha$-elements of the APOGEE SMC sample were first explored in \citet{Nidever2020}, where it was noted that the SMC also experienced very weak (low efficiency) early SF. This is shown in the O, Mg, Si, and Ca panels of Figure \ref{fig:mw_comp}, where the SMC shows a declining $\alpha$-element abundance ratio from -2.2 $<$ [Fe/H] $<$ -1.5. The [$\alpha$/Fe] abundances for the SMC over this range are $\sim$ 0.05 dex lower than the LMC [$\alpha$/Fe] abundances. When using Mg as a reference element, this slightly deficient [$\alpha$/Fe]-[Fe/H] abundance pattern manifests as a slightly enhanced [Fe/Mg]-[Mg/H] abundance pattern from -2.0 $<$ [Mg/H] $<$ -1.0.

The [Mg/Fe]-[Fe/H] abundance pattern is similar to the LMC in that there is a slight increase in [Mg/Fe] beginning at [Fe/H] $\simeq$ -1.3, with a peak at [Fe/H] $\simeq$ -1.0, followed by a slight decrease. The [O/Fe], [Si/Fe], and [Ca/Fe] abundance patterns are flat over this range. The [$\alpha$/Mg]-[Mg/H] abundance patterns are nearly identical to the LMC, with the SMC only extending to [Mg/H] $\simeq$ -0.8, $\sim$0.4 dex lower than the metal-rich extent of the LMC.

\subsubsection{Carbon and Nitrogen}

Both the [(C+N)/Fe]-[Fe/H] and [(C+N)/Mg]-[Mg/H] abundance patterns of the SMC are similar to those of the LMC. At [Fe/H] $>$ -0.8, the LMC [(C+N)/Fe] abundance begins to increase with increasing [Fe/H] whereas the SMC appears to remain flat before ending $\sim$ 0.3-0.4 dex below the MW trend. Because the [(C+N)/Mg]-[Mg/H] abundance of the SMC follows that of the LMC, the slight difference at the metal-rich end of the [(C+N)/Fe]-[Fe/H] abundance pattern is likely due to different amount of Fe from Type Ia SNe between the two galaxies.

\subsubsection{Aluminum and Nickel}

The Al and Ni abundance patterns of the SMC again are very similar to those of the LMC. The SMC shows slight deficiencies in [Al/Fe] and [Ni/Fe] as compared to the LMC at [Fe/H] $<$ -1.5, as well as at [Fe/H] $>$ -0.8. However, the [Ni/Mg] and [Al/Mg] abundance patterns show that like for the other elements, this difference is driven by a difference in Fe.

\subsubsection{Cerium}

The [Ce/Fe]-[Fe/H] and [Ce/Mg]-[Mg/H] patterns of the SMC are similar to those of the LMC, including the apparent change in increase of [Ce/Mg] abundance. In the case of the SMC, the respective slopes of the [Ce/Mg] increases are 1.6 and 0.9,
with the break in the slope occurring at [Mg/H]$\sim$-1.3.  

\subsubsection{SMC Interpretations}

The similarities between the SMC and LMC abundance patterns show that they indeed had a similar chemical evolution history, with the LMC enriching to higher metallicities than the SMC by $\sim$ 0.4 dex, potentially a consequence of its larger mass. However, there are two subtle differences between the two. First, the SMC had slightly weaker early star formation efficiency, as shown by its slightly lower $\alpha$-element abundance for the metal-poor stars (or enhanced [Fe/Mg]). Second, while the small rise of [Mg/Fe] at [Fe/H] $\simeq$ -1.0, and the change in slope of the [Ce/Mg]-[Mg/H] abundance pattern all suggest the SMC experienced an increase of Type II SNe, this increase was weaker than that in the LMC. We quantify the relative strengths and timing of these star formation events in the galaxies in \S \ref{sec:chem_results}.

\subsection{GSE}
\label{sec:results_gse}

The middle columns of Figures \ref{fig:mw_comp} and \ref{fig:mw_comp_mg} show the abundance pattern of GSE, and the median tracks are plotted in green in Figure \ref{fig:medians}. The abundance patterns of this dwarf were studied in detail before it was even confirmed to be a separate entity (e.g., \citealt{Nissen2010,Schuster2012,Hawkins2015,Fernandez-Alvar2018,Hayes2018a}). The characteristic abundance pattern of this galaxy is the declining [$\alpha$/Fe] abundance pattern with increasing [Fe/H] that occurs at lower [Fe/H] values than the MW (-1.2 $\lesssim$ [Fe/H] $\lesssim$ -0.7), resulting in a ``separate'' sequence from the MW disk sequences. There have been some recent works that have fit chemical evolution models to this abundance pattern, finding one major star formation epoch that was cut off some 10 Gyr ago, presumably by its in-fall onto the MW environment (e.g., \citealt{Fernandez-Alvar2018,Gallart2019,Vincenzo2019}).

\subsubsection{O, Mg, Si, and Ca}

The $\alpha$-element abundance patterns of the GSE look like a more metal-poor MW high-$\alpha$ disk track. The $\alpha$ elements are relatively flat at [Fe/H] $<$ -1.3, but decline at [Fe/H] $>$ -1.2, with the most metal-rich GSE stars reaching solar [$\alpha$/Fe] abundances. Compared to the other dwarf galaxies, GSE is enhanced in [$\alpha$/Fe] over the metallicity range -1.5 $<$ [Fe/H] $<$ -0.7, but still $\sim$ 0.10-0.15 dex below the MW high-$\alpha$ sequence.  The difference in Fe abundance at fixed [Mg/H] is highlighted in the [Fe/Mg]-[Mg/H] panel of Figure \ref{fig:medians}, where the GSE abundance pattern lies much closer to the MW high-$\alpha$ sequence (low-Fe) than that of the other dwarf galaxies.

The [O/Mg], [Si/Mg], and [Ca/Mg] abundances of GSE are all within $\sim$ 0.05 dex of solar over its full metallicity range, consistent with production by a population of massive stars that is similar to the MW's. Relative to the other dwarf galaxies, [O/Mg] and [Si/Mg] are nearly indistinguishable, but [Ca/Mg] is slightly depressed, perhaps because Ca has a larger Type Ia SNe contribution and the Type Ia enrichment of GSE is lower as shown by its depressed [Fe/Mg].

\subsubsection{Carbon and Nitrogen}

Unlike the MCs, the [(C+N)/Fe] abundances of the majority of GSE stars are flat across nearly the entire metallicity range, with some of the more metal-poor stars scattering to higher values of [(C+N)/Fe]. The [(C+N)/Mg]-[Mg/H] abundance pattern of GSE is a metal-poor extension of the MW high-$\alpha$ sequence (perhaps best seen in the upper-right panel of Figure \ref{fig:medians}), with the GSE abundance pattern perhaps being slightly enhanced by $\sim$ 0.05 dex, but still deficient as compared to the other dwarf galaxies.

\subsubsection{Aluminum and Nickel}

Aluminum is one of the elements that literature works have often used to select out what were then referred to as ``accreted halo'' stars from MW samples (e.g., \citealt{Hawkins2015}), and the [Al/Fe]-[Fe/H] abundance patterns of GSE presented here show that the GSE stars with [Fe/H] $>$ -1.2 are $\sim$ 0.2-0.3 dex deficient from the MW. This deficiency is shared by the other dwarf galaxies, although the [Al/Fe]-[Fe/H] abundance patterns of the dwarf galaxies differ slightly in shape from each other. However, the [Al/Mg]-[Mg/H] abundance patterns emphasize that the differences among the dwarfs are again driven by differences in Fe rather than differences in Al, as nearly all dwarf galaxies share the same [Al/Mg] abundance pattern. Their trends remain substantially below an extension of the MW [Al/Mg] trend.

Similar to the $\alpha$ elements, the [Ni/Fe]-[Fe/H] abundance pattern is slightly deficient ($\sim$ 0.05 dex from the most metal-rich stars) as compared to the MW abundance pattern, but still enhanced as compared to the other dwarf galaxies ($\sim$ 0.05 dex). The [Ni/Mg] abundance of GSE is closer to the pure Type II abundance ratio of the MW high-$\alpha$ disk, slightly below that of the other dwarf galaxies.

\subsubsection{Cerium}

The [Ce/Fe]-[Fe/H] abundance pattern of GSE is similar to the other dwarf galaxies. However, the [Ce/Mg]-[Mg/H] abundance plots show that at fixed [Mg/H], GSE is slightly deficient in [Ce/Mg] ($\sim$ 0.10-0.15 dex) as compared to the other dwarf galaxies. 

\subsubsection{GSE Interpretations}

The relatively simple $\alpha$-element abundance patterns can be explained by Type II SNe dominating at low metallicities, before Type Ia SNe exploding after some time delay, diluting the $\alpha$ elements as Fe is added to the ISM in large quantities (e.g., \citealt{Tinsley1979}). Whatever extended star formation GSE might have had was cutoff by its proximity to the MW. As such, we see no sign of flat or increasing $\alpha$-element abundance patterns suggesting a starburst like the MCs, or any extended SFH. However, the fact that GSE was $\alpha$-element enhanced until [Fe/H] $\sim$ -1.2 shows that the early SF of this galaxy was much more efficient than the early SF of the MCs, which were $\alpha$-element enhanced until [Fe/H] $\sim$ -2.2. This difference in star formation history also results in slightly enhanced [Al/Fe]-[Fe/H] and [Ni/Fe]-[Fe/H] abundance patterns as compared to the other dwarf galaxies, but these differences are largely driven by the ``extra'' Fe from the increased relative contribution from Type Ia SNe in the other dwarf galaxies.

The slight deficiencies observed in [Ce/Mg] and [(C+N)/Mg] observed for GSE suggest that GSE had slightly lower contributions from AGB stars to its evolution as compared to these galaxies. This is expected if GSE only evolved over the course of 2-4 Gyr, before large amounts of AGB stars could contribute to its evolution.

\subsection{Sgr}
\label{sec:results_sgr}

The chemistry of the Sgr dwarf has been studied by numerous authors (e.g., \citealt{Chou2007,Sbordone2007,McWilliam2013,Hasselquist2017,Carlin2018,Hansen2018}), with \citet{Hayes2020} analyzing the largest sample of core and stream stars using APOGEE DR16. In general, these analyses all find the [X/Fe] abundances of the more metal-rich stars in the Sgr core to be below the MW abundance trends. Interpretations of such sub-solar abundance ratios range from high Type Ia/Type II SNe ratio to top-light IMF. Here we analyze a sample of stars that is essentially an expanded sample of \citet{Hayes2020}. 

While a detailed analysis of the spatial dependence of the chemical abundance patterns of Sgr is beyond the scope of this work, we find that the Sgr stream sample is $\sim$ 0.5 dex more metal-poor than the main body sample (see e.g., \citealt{Hayes2020}). However, we verify that stars with similar metallicities between the two samples have near-identical chemical abundance patterns (see \S \ref{app:targ_sgr} for more details).

\subsubsection{O, Mg, Si, and Ca}

The $\alpha$-elements in Sgr smoothly decline from -2.5 $<$ [Fe/H] $<$ -0.9, going from the MW ``halo'' high-$\alpha$ plateau at the metal-poor end to below the MW low-$\alpha$ ``thin'' disk trend at [Fe/H] = -0.9. Sgr begins this decline at a slightly higher metallicity than the MCs, but a lower metallicity than GSE. At [Fe/H] $>$ -0.9, the [O/Fe], [Si/Fe] and [Ca/Fe] abundances are nearly flat at solar or slightly sub-solar values, but the [Mg/Fe] abundance shows a slight increase followed by a decrease, as also observed in the SMC. The [$\alpha$/Fe] abundance of Sgr at [Fe/H] $>$ -0.9 remains $\sim$ 0.1 dex below the LMC trend and MW low-$\alpha$ sequence.

While Sgr extends to slightly higher values of [Fe/H] than the LMC, both galaxies enriched to nearly the same level of [Mg/H], with the Sgr abundance trend being deficient in [Fe/Mg] as compared to the MCs at [Fe/H] $>$ -1.0, but enhanced at [Fe/H] $>$ -1.0. The Sgr stars with [Mg/H] $>$ -0.5 are very slightly enhanced in [Ca/Mg] as compared to the LMC, but otherwise the [Ca/Mg], [O/Mg], and [Si/Mg] patterns are nearly identical to those of the LMC.

\subsubsection{Carbon and Nitrogen}

The [(C+N)/Fe]-[Fe/H] abundance pattern of Sgr is similar to that of the LMC. The increase in [(C+N)/Fe] occurs at [Fe/H] $\simeq$ -0.8 for both galaxies, but the slope of the increase with [Fe/H] is shallower in Sgr than the LMC, so the sequences diverge at higher [Fe/H].

At [Mg/H] $>$ -0.7, the Sgr [(C+N)/Mg] abundance trend is $\sim$ 0.05 dex above the LMC trend, with the most metal-rich Sgr stars intersecting the MW low-$\alpha$ trend, showing that at these metallicities Sgr formed stars with an AGB/Type II SNe ratio that was closer to that of the MW low-$\alpha$ disk.

\subsubsection{Aluminum and Nickel}

The [Al/Fe] abundance pattern of Sgr is $\sim$ 0.4 dex below the MW trend at [Fe/H] $>$ -0.8. This is even more deficient than the other galaxies mentioned thus far. However, as shown in the [Al/Mg] abundance plane in Figure \ref{fig:mw_comp_mg} and in Figure \ref{fig:medians}, the Sgr [Al/Mg]-[Mg/H] abundance pattern is nearly indistinguishable from the other dwarf galaxies, with both the LMC and Sgr remaining noticeably deficient ($\sim$ 0.2 dex) compare to the MW in [Al/Mg] at [Mg/H] $>$ -0.7. 

The [Ni/Fe]-[Fe/H] pattern of Sgr largely follows that of the LMC, but it continues to decline at [Fe/H] $>$ -0.9, at which point the LMC [Ni/Fe]-[Fe/H] abundance patterns becomes flat/slightly increasing. The [Ni/Mg]-[Mg/H] pattern of Sgr at [Mg/H] $>$ -0.8 lies in between the MW low-$\alpha$ sequence and the LMC. The [Ni/Mg] differences follow the [Fe/Mg] differences, which suggests that they are driven by differing levels of Type Ia SNe enrichment. 

\subsubsection{Cerium}

The [Ce/Fe]-[Fe/H] abundance pattern of Sgr is similar to those of the other galaxies, with the most metal-rich Sgr stars being enhanced in [Ce/Fe] as compared to the MW stars. The median tracks in Figure \ref{fig:medians} show that the most metal-rich Sgr stars are slightly enhanced in [Ce/Mg] as compared to the LMC, and moderately enhanced ($\sim$ 0.2 dex) as compared to the MW.

\subsubsection{Sgr Interpretations}

The $\alpha$-element abundance patterns show that early on in its evolution, Sgr experienced relatively weak star formation as compared to the MW and GSE, but stronger than the MCs and Fnx. Sgr then evolved to a much higher Type Ia/Type II SNe ratio than the other dwarf galaxies, with a [Fe/Mg] ratio that is enhanced over the MCs and MW low-$\alpha$ sequence at [Mg/H] $>$ -1.0. Despite this clear difference in early star formation efficiency, Sgr and LMC enrich to nearly the same levels of [Fe/H], with Sgr extending to metallicities that are $\sim$ 0.2 dex higher than the LMC. The increased early SFE and enhanced enrichment is unexpected in the paradigm of the mass-metallicity relationship \citep{Kirby2011}, as the Sgr dwarf was thought to be much less massive than the LMC. However, the two galaxies do enrich to the same level of [Mg/H], implying the tension is somewhat reduced if [Mg/H] is used to track metallicity instead of [Fe/H]. To more accurately analyze where these two galaxies lie on the mass-metallicity relationship, we would need to better account for selection biases, which is beyond the scope of this work. Both the final metallicity and the early SFE seem to show Sgr behaving as though it were a fairly massive dwarf galaxy.

The flat, or near-flat, in the case of Mg, [$\alpha$/Fe]-[Fe/H] abundance patterns at [Fe/H] $>$ -0.9 imply that Sgr did experience some extended SF, with an increase in Mg from Type II SNe preventing the [Fe/Mg] abundance from rising further. This extended star formation event could have been started as Sgr began falling into the MW. As shown in \citet{Hayes2020}, the Sgr stream does not contain stars with [Fe/H] $\gtrsim$ -0.50, but the Sgr [$\alpha$/Fe] abundance becomes flat with increasing [Fe/H] at [Fe/H] $\sim$ -0.9. So the extended star formation occurred before some stars were stripped as well as after, plausibly corresponding with pericenter passages through the disk of the MW (see also \citealt{Ruiz-Lara2020}).

The low [Ni/Fe] in Sgr was interpreted as evidence for a top-light IMF in some literature works (e.g., \citealt{Hasselquist2017}). However, analyses of MW median abundance trends implies that Type II SNe contribute a larger fraction of Ni than of Fe \citep{Weinberg2019}, so this deficiency could also arise from differences in the Type Ia/Type II SNe ratios. Both the [Ni/Fe] and [Al/Fe] abundance patterns can largely be attributed to Sgr evolving to a higher Ia/II ratio than the MW and MCs, increasing the amount of Fe in Sgr. However, Sgr shares the peculiar deficiency relative to the MW in [Al/Mg] at [Mg/H] $>$ -0.5, which is plausibly explained by more metal-poor Type II SNe contributing to the enrichment than the metallicity of those SNe that contributed to the Al enrichment in Sgr, perhaps because of its lower SFE.

The [(C+N)/Mg] and [Ce/Mg] abundance patterns imply that Sgr had more contribution from AGB stars to its enrichment than the LMC and the MW. This is perhaps an expectation given the higher Type Ia/Type II SNe ratio in Sgr, as implied by the other abundance patterns, allowing for a stronger AGB contribution as well.

\subsection{Fnx}
\label{sec:results_fnx}

The abundances of Fnx have been studied in a variety of literature studies, some of which are comparable in number to what we have in APOGEE. In general, these studies find that Fnx exhibits a relatively metal-poor $\alpha$-element abundance ``knee'' (e.g., \citealt{Hendricks2014}), with sub-MW [$\alpha$/Fe] abundance ratios at [Fe/H] $>$ -1.5 (e.g., \citealt{Letarte2010,Hendricks2014}). These studies suggest Fnx underwent some extended SFH and formed stars from gas with a larger Type Ia/Type II SNe ratio than the MW at similar metallicity. The APOGEE data, shown in the right columns of Figure \ref{fig:mw_comp} and \ref{fig:mw_comp_mg}, contains Fnx stars with -2.2 $<$ [Fe/H] $<$ -0.5. However, because the Fnx sample is generally lower in S/N as compared to the other galaxies, most of our analysis is focused on the Fnx stars with [Fe/H] $> -1.2$, as the stars more metal poor than this have larger abundance uncertainties. The APOGEE Fnx sample is also relatively sparse at low-metallicity, consisting of $\sim$20 stars with -2.2 $<$ [Fe/H] $<$ -1.2.

\subsubsection{O, Mg, Si, and Ca}

While it is difficult to analyze the metal-poor ``knee'' of Fnx using this APOGEE sample, the $\alpha$-element abundances decline from -2.2 $<$ [Fe/H] $<$ -1.2. This is a similar range of decrease as the MCs, although Fnx reaches a much more deficient [$\alpha$/Fe] abundance  at [Fe/H] = -1.2 than the MCs ($\sim$ 0.2 dex deficient). At [Fe/H] $>$ -1.2, there is a slight increase in the $\alpha$-element abundances, with the most metal-rich Fnx stars reaching the same [$\alpha$/Fe] abundance of the Sgr stars at [Fe/H] $>$ -0.8. 

The exceptionally low [Mg/Fe]-[Fe/H] manifests as an exceptionally enhanced [Fe/Mg]-[Mg/H] abundance pattern, where for fixed [Mg/H], Fnx is $\sim$0.2 dex enhanced in Fe from the MCs, $\sim$0.3 dex enhanced from GSE, and $\sim$0.4 dex enhanced over the MW high-$\alpha$ sequence. Fnx is also enhanced in [Ca/Mg] relative to the other galaxies at [Mg/H] $<$ -1.0, but is nearly identical to the other galaxies in [O/Mg] and [Si/Mg].

\subsubsection{Carbon and Nitrogen}

Like the other galaxies, Fnx shows a slight increase in [(C+N)/Fe] with [Fe/H] at [Fe/H] $>$ -1.0, but even the most metal-rich Fnx stars remain $\sim$ 0.3 dex below the MW trend. At fixed [Mg/H], Fnx is the most enhanced galaxy in [(C+N)/Mg], with an abundance that is $\sim$ 0.1 dex above the MCs and Sgr over the range -1.5 $<$ [Mg/H] $<$ -1.0.

\subsubsection{Aluminum and Nickel}

Despite having the most deficient [Al/Fe]-[Fe/H] pattern of all of the galaxies, the Fnx [Al/Mg] is nearly identical to the other dwarf galaxies. Fnx is also most deficient in [Ni/Fe], with [Ni/Fe] $\simeq$ -0.1 dex across much of the metallicity range, $\sim$ 0.1 dex below the MCs and Sgr, although the most metal-rich Sgr stars reach nearly this low [Ni/Fe] abundance. The [Ni/Mg] abundance of Fnx is enhanced compared to GSE and the high-$\alpha$ MW sequence, but about the same as the MCs and Sgr, at least at [Mg/H] $>$ -1.5.

\subsubsection{Cerium}

The [Ce/Fe]-[Fe/H] abundance pattern of Fnx largely traces those of the other galaxies, although Ce is only measurable for $\sim$ 40\% of the Fnx sample. At -1.4 $<$ [Mg/H] $<$ -1.0, Fornax is 0.2-0.4 dex enhanced in [Ce/Mg] relative to the other dwarf galaxies, and 0.5 dex enhanced relative to the MW high-$\alpha$ sequence. This level of enhancement in Fnx at [Mg/H] = -1.0 is similar to that of the MW low-$\alpha$ sequence at [Mg/H] $\sim$ -0.2.

\subsubsection{Fnx Interpretations}

Fnx is the most striking outlier among the five dwarf galaxies, in [O/Fe], [Mg/Fe], [Si/Fe], [Ni/Fe], [(C+N)/Mg], [Ca/Mg], and to a lesser extend, [Ce/Mg]. Possibly these differences could imply different IMF-averaged yields from Fnx stars, e.g., Type II SNe that produced more Fe relative to $\alpha$-elements. Another possibility is winds that preferentially ejected Type II SNe products. Or, Fnx could simply have a SFH that led to an exceptionally high ratio of Type Ia/Type II SNe enrichment. A high relative contribution of Type Ia SNe could explain most of these anomalies, and a similarly high contribution of AGB enrichment could explain high [(C+N)/Mg] and [Ce/Mg] ratios.

\subsection{[C/N] as an Age Indicator}
\label{sec:cn_age}

As has now been shown in many works (e.g., \citealt{Masseron&Gilmore2015,Martig2016a,Ness2016b,Hasselquist2019b}), we can use the APOGEE [C/N] abundance ratio as a mass/age indicator for APOGEE red giant stars. More massive stars mix a larger amount of CNO-processed material into their convective envelopes when ascending the red giant branch, which lowers their [C/N] abundance ratio because C is depleted and N is enhanced during CNO burning. However, the [C/N] abundance is only a reliable mass indicator at [Fe/H] $\gtrsim$ -0.6, below which some extra mixing occurs to further alter the [C/N] abundance (see e.g., \citealt{Shetrone2019}). We therefore limit the following [C/N]-age analysis to these metallicities, meaning that we only study the [C/N] abundance patterns of the more metal-rich LMC and Sgr stars.

\begin{figure}
\includegraphics[width=1.0\hsize,angle=0]{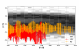}
\caption{[C/N] as compared to the MW just for Sgr (orange) and LMC (red). Running medians of 30 stars per bin are shown for the dwarf galaxies, along with the standard deviation in each bin. The MW sample is plotted as a grayscale density image.}
\label{fig:mw_comp_cn}
\end{figure}

In Figure \ref{fig:mw_comp_cn} we show how the [C/N] abundance tracks of these two galaxies as compared to those of the MW, again selecting a MW comparison sample of 3600 K $<$ T$_{\rm eff}$ $<$ 4200 K to remove the effects of potential T$_{\rm eff}$ systematic uncertainties on abundance determination. As was done in Figure \ref{fig:medians}, we calculate medians in bins of 30 stars, and plot those medians as well as the standard deviations.

From Figure \ref{fig:mw_comp_cn} we see that the Sgr stars show a [C/N] abundance pattern that falls in the same region of this abundance space as the MW low-$\alpha$ or ``thin'' disk stars ([C/N] $\simeq$ -0.3). This suggests a median age of the stars at these [Fe/H] of 3-6 Gyr (e.g., \citealt{Bensby2014,Martig2016a,Ness2016b}). This age estimate of the metal-rich Sgr stars is qualitatively consistent with the ``intermediate age metal-rich population'' described in \citet{Alfaro-Cuello2019}, who found that the stars in their sample corresponding to metal-rich Sgr stars had a mean metallicity of [Fe/H] = 0.29 $\pm$ 0.16 and mean age of 4.28 $\pm$ 1.16 Gyr. The LMC stars, on the other hand, have median [C/N] abundances that are below the Sgr trend by 0.2-0.3 dex across the entire metallicity range. This implies an age of these stars of $<$ 3 Gyr. However, the larger standard deviation in these bins implies an age range at fixed metallicity, one that is potentially spatially dependent.

While we do not explicitly map to age in this work, \citet{Hasselquist2020} derived ages for a few of these metal-rich stars and found a median age of the LMC stars of 2 Gyr, and a median age of the Sgr stars of 5 Gyr. Additionally, Povick et al. in prep  derive ages for these stars and find that most stars with [Fe/H] $>$ -0.5 are younger than 2 Gyr. This is consistent with the results of the $\alpha$-element abundances presented above and in \citet{Nidever2020}, which highlight that the metal-rich LMC stars likely formed in a recent burst of star formation. This picture is also consistent with literature studies of the SFH of the LMC (e.g., \citealt{Harris&Zaritsky2009,Weisz2013,Monteagudo2018,Ruiz-Lara2020b,Nidever2021}). Our interpretation presumes that the birth [C/N] ratios of the metal-rich Sgr, LMC, and low-$\alpha$ MW sequence are similar such that these differences in [C/N] reflect differences in dredge-up materials that in turn reflect differences in stellar mass.

\subsection{Chemical Abundance Summary}

The $\alpha$ elements, O, Mg, Si, and Ca, show that out of all these dwarf galaxies, GSE had the highest early star formation efficiency, followed by Sgr, then the MCs. Our Fnx sample does not contain enough metal-poor stars to precisely place its early star formation efficiency in comparison to the other galaxies, but we do find that Fnx evolved to the highest Type Ia/Type II SNe ratio, with a [Mg/Fe] abundance that is $\sim$0.15 dex below the MCs and Sgr at [Fe/H] = -1.2.  All galaxies except for GSE show a flattening or even increase in their $\alpha$-element abundance pattern ([$\alpha$/Fe]), suggesting an extended star formation event that polluted the ISM with many more Type II SNe. The LMC had the strongest second star formation event, which actually slightly enhanced the [$\alpha$/Fe] abundance patterns. The [C/N] abundance patterns suggest that this strong second star formation event occurred at more recent times than the second star formation epoch in Sgr.

All of the dwarf galaxies are deficient in [Ni/Fe] to some level as compared to the MW, with GSE being slightly deficient ($\sim$ 0.1 dex at most) and Fnx being the most deficient ($\sim$ 0.35 dex at most). However, the fact that the [Ni/Mg] abundances of these galaxies are below the MW low-$\alpha$ sequence suggests that there is some metallicity dependence on the production of Ni, since the enhanced Type Ia/Type II SNe ratio implied by the $\alpha$-element abundance pattern does not not result in an enhanced [Ni/Mg] abundance. Another plausible explanation of deficient [Ni/Fe] is the contribution of sub-Chandrasekhar Type Ia SNe to the chemical enrichment of a galaxy (e.g., \citealt{McWilliam2018,Hill2019,Kirby2019}), although \citet{Kirby2019} finds minimal evidence for this in the dwarf galaxies that had extended star formation histories (e.g., Fornax), favoring the metallicity-dependent production of Ni as the explanation for the abundance patterns presented here.

More metal-poor SNe is also a plausible explanation for the deficient [Al/Mg] abundances of the dwarf galaxies, with the stars forming at [Mg/H] $>$ -0.7 forming from gas that was polluted preferentially bye lower metallicity SNe than the MW sequences, which evolved much quicker.

The [(C+N)/Mg] and [Ce/Mg] abundance patterns show that all dwarf galaxies had higher contributions to their evolution from AGB stars than the MW high-$\alpha$ sequence. Compared to each other over the metallicity range -1.5, $<$ [Mg/H] $<$ -1.0, GSE had the weakest contribution from AGB stars, followed by Sgr, then the MCs, and finally Fnx. Only the LMC and Sgr evolved substantially past [Mg/H] = -1.0, at which point the relative AGB contribution to their chemical enrichment diverged, with the LMC experiencing an increase in Mg from the strong starburst.

From these chemical abundance comparisons, we find evidence that galaxies in denser environments undergo high-efficiency formation events early on in their histories. Conversely, more isolated galaxies have very low-efficiency star formation events early on, and form apparently large fractions of their stars as a result of interactions: Sgr with the MW (e.g., \citealt{Ruiz-Lara2020}), MCs interacting with each other, and Fnx with a small merger (e.g., \citealt{Coleman2005,Leung2020} ).

\section{Chemical Evolution Modeling}
\label{sec:mod}

In this section, we implement chemical evolution models to infer physical parameters of the star formation histories of these galaxies. Note that we are using these tools to estimate the star formation histories of these galaxies using the APOGEE chemistry alone. This allows us to quantify the relative strengths of the starbursts of the LMC and SMC, for example. While we do compare these star formation histories to photometric studies in \S \ref{sec:photo_comp}, a much more quantitative and robust analysis of these data can be done in which the photometry is \textit{combined} with the spectroscopic abundance data, resulting in much tighter constraints of the star formation histories. This analysis is beyond the scope of this work.

To extract parameters of the SFHs of each galaxy, we adopt two chemical evolution codes: flexCE \citep{Andrews2017} and the model described in \citet{Lian2018a} and \citet{Lian2020a}, hereafter referred to as ``the Lian model''.

While the models are similar at a fundamental level, there are several distinct differences (explained in detail in Appendix B) that make the inclusion of both of them valuable. First, the flexCE model was tuned in its yields and fiducial parameters to match the APOGEE [O/Fe]-[Fe/H] abundance pattern (which tracks very closely to [Si/Fe]; see \citealt{Andrews2017}). O is an $\alpha$-element with nucleosynthetic yields that are thought to be well understood (e.g., \citealt{Kobayashi2006}). However, the APOGEE O abundances are not as well-measured across the HR diagram as Si abundances are. Therefore, we only fit flexCE model tracks to [Si/Fe]-[Fe/H] abundances. The Lian model, on the other hand, has been tuned to fit the APOGEE [Mg/Fe]-[Fe/H] abundance plane. 

The model codes are also parameterized differently, which we describe both below and in \S \ref{app:modeling}. Results of the modeled star formation histories are shown and discussed in \S \ref{sec:chem_results}. In all cases where we consider bursts of star formation, we implement them as time-localized changes to the star formation efficiency, with a smooth gas accretion history. The effects of efficiency-driven starbursts and accretion-driven starbursts on chemical evolution are somewhat different (see e.g., \citealt{Johnson2020}). For satellite galaxies, efficiency-driven bursts caused by dynamical interactions seem the somewhat more natural choice, as the smaller gravitational potential of lower-mass satellite galaxies reduces their chance of accretion-driven starbursts. 

\subsection{FlexCE Modeling}
\label{sec:flexce}

FlexCE is a one-zone, open-box, chemical evolution modeling code that is broadly described in \citet{Andrews2016}, which also provides a fiducial model designed to fit the MW's [O/Fe]-[Fe/H] chemical abundance pattern for stars in the solar neighborhood.  Naturally dwarf galaxies are expected to have different SFHs than the MW, so many of these parameters must be changed to produce appropriate dwarf galaxy chemical evolution models.  However, we have kept some of the fiducial parameters, because they are thought to vary less from galaxy to galaxy or to be driven by stellar evolution rather than galactic evolution.  

For example, we retain the fiducial parameters for the chemical abundance yields, Type Ia supernova delay-time distribution \citep[which appears to be constant across massive galaxies at least,][]{walcher2016}, and the initial mass function (IMF). \citet{Andrews2016} use a Kroupa IMF \citep{Kroupa2001} for their MW model. While some past studies have speculated that individual dwarf galaxies may have had a more top-light IMF \citep[e.g.,][]{Carlin2018}, other studies have refuted these claims when observing lower metallicity stars \citep{Hansen2018}.  Any parameters not mentioned below or given in Tables \ref{flexce_fixed} and \ref{flexce_grid} use the fiducial values from \citet{Andrews2016}.

There are two general modifications we make to the fiducial flexCE model.  1) We use a delayed tau model for the gas inflow of our chemical evolution models \citep[motivated by cosmological simulation; e.g.,][]{Simha2014}, allowing the gas inflow to ramp up at early times in the universe before peaking and falling off at later times.  2) We add a formulation for a time variable SFE to flexCE, that scales the SFE up or down in a Gaussian shaped deviation from a constant SFE.  This addition allows us to temporarily cut off star formation or produce a burst of star formation in our models.  More information about these modifications and how they were implemented can be found in Section \ref{flexce:inflow} and \ref{flexce:sfe} for the inflow and SFE, respectively.

With these modifications we model the chemical abundances of our five galaxies.  Because there are many parameters that can produce variation in chemical evolution models, we have chosen to vary only the parameters that most strongly impact the chemistry (other than yields), and we use slightly different strategies for different galaxies.  For simplicity of modeling each system we limit the number of variable parameters to four and fix the remaining parameters at the values shown in Table \ref{flexce_fixed}.  

For each system we allow the SFE and outflow mass loading factor to differ, which control much of the global chemical abundance patterns that are produced.  The flexCE outflow parameterization assumes that enriched ISM gas is ejected proportionally to the the star formation rate, with the mass loading factor, $\eta$, as the constant of proportionality \citep{Andrews2016}.  This is a relatively simple form for outflow, but allows the model to generally track gas outflows due to stellar sources, such as stellar winds, supernovae, stellar radiation pressure, etc. In addition to these parameters we add a time variable SFE for all galaxies except GSE, since we don't see any complex features in its chemistry.  

For the LMC and SMC, because we see a bump or rise in the chemical abundance patterns that we believe is due to a recent burst of star formation, we fit these systems by adding an increase in star formation that has a variable timing and strength, with a fixed duration of $\sigma_{b} = 750$ Myr, roughly the duration of the increase in SFR modeled for the LMC in \citet{Nidever2020}.  

Some literature works found that Sgr and Fnx have had a generally bursty star formation history, with periods of star formation broken up by periods where there is very little star formation (see e.g., \citealt{Siegel2011} for Sgr and e.g., \citealt{Hendricks2014} for Fnx).  To reproduce this type of variable star formation we model these systems with a decrease in SFE to 1\% of its baseline value, with the time and duration of this decrease as free parameters.

For all five galaxies we fix the initial gas mass and inflow mass scale and use the same values across all galaxies.  While these galaxies do not have the same masses, the overall mass scale (i.e., total mass) does not impact the chemistry, and only the ratio of initial-to-inflow mass impacts the chemistry.  However, the initial gas mass has a minor effect that is largely degenerate with other parameters we vary, so we have held this ratio constant for all galaxies.  

We also fix the inflow timescale of each galaxies' model, which has some impact on the chemistry, particularly the shape of the $\alpha$-knee, but we consider this a secondary effect, and defer it to future study.  For GSE, the SMC, and the LMC, the timescale has been fixed to a value that roughly fits the shape of the $\alpha$-knee and is early enough so that most of the accretion happens before the range of starburst timings probed in the SMC and LMC. Similarly for Sgr and Fnx we use a slightly earlier inflow timescale so that most gas is accreted before the range of timescales for the stalls in star formation that we fit.

Then for each galaxy, we produce a grid of chemical evolution models with the variable parameters mentioned above.  We initially test a broad ranges of parameters before narrowing to the final grids used here.  At a minimum we require five grid steps in each dimension, but have expanded the initial grids in dimensions where the best fit value lay near a grid edge to confirm the validity of the best fit results.  The values of the grid points and their spacings are listed in Table \ref{flexce_grid}, and parameters that are not varied are marked by ``$-$'' with values listed in Table \ref{flexce_fixed}.

\begin{deluxetable*}{l c c c c c}
\tabletypesize{\scriptsize}
\tablewidth{0pt}
\tablecolumns{3}
\tablecaption{flexCE Fixed Parameters \label{flexce_fixed} (\S \ref{sec:flexce} and \S \ref{app:modeling})}
\tablehead{\colhead{System} & \colhead{Initial Gas Mass} & \colhead{Inflow Mass Scale} & \colhead{Inflow Timescale} & \colhead{Burst Duration} & \colhead{Burst Strength} \\ \colhead{Name} & \colhead{${\rm M}_0$ (${\rm M}_{\odot}$)} & \colhead{${\rm M}_{\rm i}$ (${\rm M}_{\odot}$)} & \colhead{$\tau_{\rm i}$ (Gyr)} & \colhead{$\sigma_{\rm b}$ (Gyr)} & \colhead{$\rm F_{\rm b}$}}
\startdata
LMC & $3 \times 10^{9}$ & $6 \times 10^{10}$ & 2 & 0.75 & $-$ \\
SMC & $3 \times 10^{9}$ & $6 \times 10^{10}$ & 2 & 0.75 & $-$ \\
Sgr & $3 \times 10^{9}$ & $6 \times 10^{10}$ & 1 & $-$ & 0.01 \\
GSE & $3 \times 10^{9}$ & $6 \times 10^{10}$ & 2.5 & $-$ & $-$ \\
Fnx & $3 \times 10^{9}$ & $6 \times 10^{10}$ & 1 & $-$ & 0.01
\enddata
\end{deluxetable*}

To fit our models, we compute the $\chi^2$ of each model's resulting [Si/Fe]-[Fe/H] track relative to the data for each system. As previously mentioned, Si is used for this fitting, because it is both precisely measured by APOGEE \citep{Jonsson2020} and has well understood nucleosynthetic yields that can match MW chemical abundance patterns \citep{Andrews2016}. To obtain the $\chi^2$ of each model we first calculate the model $\chi^2$ using the distance of each observed star from the model track in [Si/Fe]-[Fe/H] space and the magnitude of the star's [Si/Fe] and [Fe/H] uncertainties in that direction. We then also penalize models that evolve past the maximum metallicity of each system with an extra term to the model $\chi^2$ that is the distance of each model time step from the high metallicity end of each galaxy scaled by the typical abundance uncertainties at those metallcities.  We only turn off this penalization for GSE because we know that its star formation was cutoff after some time (which we use as a check on our chemical evolution model for GSE), and for the last 1 Gyr of evolution in the remaining galaxies, because we wouldn't expect to observe RGB stars of such young age.

To find the optimal solution for each of our fit parameters, we then take the model with the minimum $\chi^2$ as our best fit.   The best fit values of each parameter can be found in Table \ref{flexce_grid} for our five galaxies.

\begin{deluxetable*}{l c c c c c}
\tabletypesize{\scriptsize}
\tablewidth{0pt}
\tablecolumns{3}
\tablecaption{flexCE Variable Parameters' Model Grid Ranges \label{flexce_grid} (\S \ref{sec:flexce} and \S \ref{app:modeling})}
\tablehead{\colhead{System} & \colhead{SFE} & \colhead{Outflow} & \colhead{Burst Time} & \colhead{Burst Duration} & \colhead{Burst Strength} \\ \colhead{Name} & \colhead{(Gyr$^{-1}$)} & \colhead{$\eta_{\rm wind}$} & \colhead{$\tau_{\rm b}$ (Gyr)} & \colhead{$\sigma_{\rm b}$ (Gyr)} & \colhead{$\rm F_{\rm b}$}}
\startdata
\cutinhead{Model Grid}
LMC & 0.01$-$0.03 ($\Delta = 0.005$)& 2$-$10 ($\Delta = 2$) & 8$-$12 ($\Delta = 1$) & $-$ & 4$-$8 ($\Delta = 1$) \\
SMC & 0.006$-$0.01 ($\Delta = 0.001$)& 5$-$25 ($\Delta = 5$) & 6$-$11 ($\Delta = 1$) & $-$ & 2$-$10 ($\Delta = 2$) \\
Sgr & 0.02$-$0.06 ($\Delta = 0.01$)& 12.5$-$22.5 ($\Delta = 2.5$) & 4$-$8 ($\Delta = 1$) & 0.25$-$1.25 ($\Delta = 0.25$) & $-$ \\
GSE & 0.08$-$0.2 ($\Delta = 0.01$) & 1$-$11 ($\Delta = 1$) & $-$ & $-$ & $-$ \\
Fornax & 0.01$-$0.05 ($\Delta = 0.01$)& 20$-$120 ($\Delta = 20$) & 2$-$8 ($\Delta = 1$) & 1$-$6 ($\Delta = 1$) & $-$ \\
\cutinhead{Best Fit}
LMC & 0.015 & $\ \, \ \, 2$ & 11 & $-$ & 6 \\
SMC & 0.008 & $\ \, 10$ & $\ \, 7$ & $-$ & 4 \\
Sgr & $0.03 \ \,$ & $\ \, \ \, \ 17.5$ & $\ \, 5$ & $\ \, \ 0.5$ & $-$ \\
GSE & $0.14 \ \,$ & $\ \, \ \, 6$ & $\ \, -$ & $-$ & $-$ \\
Fornax & $0.03 \ \,$ & 100 & $\ \, 6$ & 5 & $-$ 
\enddata
\end{deluxetable*}

\subsection{Lian Modeling}
\label{sec:lian}

The Lian chemical evolution model is a one-zone open-box model that considers the metal production and depletion by star formation, gas accretion and galactic winds. The star formation rate is determined from the gas mass following the form of the Kennicutt-Schmidt star formation law (SFL, \citealt{Kennicutt1998}), assuming a Kroupa IMF \citep{Kroupa2001}. The SFE is thus regulated by the coefficient of the SFL; 
we assume a constant coefficient ($C_{\rm initial}$) unless a starburst event occurs. A different version of this model with more flexible gas accretion and star formation histories has been successfully applied to interpret the stellar chemistry observations in various components of the Milky Way, including the bulge \citep{Lian2020c}, inner disk \citep{Lian2020b}, and outer disk \citep{Lian2020a}. For more details about the nucleosynthesis prescription and development of the basic model, we refer the reader to \citet{Lian2018a} and \citet{Lian2020a}.

We include here two major modifications of the \citet{Lian2018a} and \citet{Lian2020a} models. First, the gas inflow parameterization is simpler, with our approach assuming gas accretion to decline exponentially (compare to \S \ref{sec:flexce}): $A(t)=A_{\rm initial}e^{-t/\tau_{\rm acc}}$, where $A_{\rm initial}$ is the initial gas accretion rate and $\tau_{\rm acc}$ is the declining timescale.  In this way, the burst event is described by three parameters: the timescale ($\tau_{\rm burst}$), start time ($t_{\rm start}$) and duration ($\Delta t$) of the SFE increase. After the burst, the coefficient of the SFL is set to decrease exponentially. Since this paper mainly focuses on the burst event, for simplicity, we fix this decreasing timescale to be 0.2 Gyr. 

The outflow is characterized as in flexCE, with the strength regulated by the mass loading factor, $\lambda$. As stronger outflows remove more gas from the galaxy, to keep the present-day stellar mass predicted by various models fixed, we set the initial gas accretion rate to scale with the outflow strength, i.e. $A_{\rm initial}\propto(1+\lambda)$. 

We have six free parameters in total: two parameters characterizing the initial gas accretion and star formation histories ($\tau_{\rm acc}$ and $C_{\rm initial}$), three parameters describing the starburst event ($\tau_{\rm burst}$, $t_{\rm start}$, and $\Delta t$), and one parameter determining the strength of gas outflow ($\lambda$), shown in Table \ref{lian_grid}.

We build a grid of models with each chemical evolution code and find the best-fit model through chi-squared minimization. As in flexCE, we fit only to the chemical evolution tracks and not to the density.

\begin{deluxetable*}{l c c c c c c}
\tabletypesize{\scriptsize}
\tablecolumns{7}
\tablecaption{Lian Variable Parameters' Model Grid Ranges \label{lian_grid} (\S \ref{sec:lian} and \S \ref{app:lian})}
\tablehead{\colhead{System} & \multicolumn{2}{c}{Gas Inflow} & \colhead{Outflow} & \multicolumn{3}{c}{Star Burst}  \\ \colhead{Name} & \colhead{$\tau_{acc}$} & \colhead{C$_{\rm initial}$} & \colhead{$\lambda_{\rm wind}$} & \colhead{$\tau_{\rm burst}$} & \colhead{$t_{\rm start}$} & \colhead{$\Delta$ t}} 
\startdata
\cutinhead{Grid}
 \ & 2, 10 & 0.001$-$0.1($\Delta$log = 0.5) & 0$-$20($\Delta$ = 5) & -0.5$-$-8($\Delta$log = 0.3) & 4$-$12($\Delta$ = 2) & 1$-$6($\Delta$=1) \\
\cutinhead{Best Fit}
LMC & 10 & 0.01 & $\ \, 15$ & -2  & $\ \, 10$ & 3\\
SMC & 10 & 0.01 & \, 20 & \ \, -1 & $\ \, 8$ & 1 \\
Sgr & 2 & \, $0.01 \ \,$ & \, 10 &  \ \, -4 & \, 6 & 4\\
GSE & 2 & \, $0.01 \ \,$ & $\ \, 5 $ &  \ \, -2 & \, - & 2 \\
Fornax & 2 & \, 0.032 & \, 20 & -1 & \, 6 & 1 
\enddata
\end{deluxetable*}

\subsection{Chemical Evolution Modeling Results}
\label{sec:chem_results}

\begin{figure*}[t]
\includegraphics[width=1.0\hsize,angle=0]{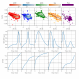}
\caption{Top row: [Si/Fe]-[Fe/H] abundance space for each of the five dwarf galaxies, along with their best-fit chemical evolution track as inferred from the flexCE code \citep{Andrews2017} over-plotted in black. Middle row: star formation histories of each galaxy, normalized by the peak star formation rate. Vertical dotted lines indicate the time of peak SF. Bottom row: metallicity evolution with time for each dwarf galaxy. Horizontal dotted lines indicate the [Fe/H] at the peak of the SFH.}
\label{fig:flexce}
\end{figure*}

\begin{figure*}[t]
\includegraphics[width=1.0\hsize,angle=0]{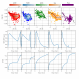}
\caption{Same as in Figure \ref{fig:flexce}, but with the Lian models that are fit to [Mg/Fe]-[Fe/H] instead of [Si/Fe]-[Fe/H].}
\label{fig:lian}
\end{figure*}

Chemical evolution modeling results are shown in Figure \ref{fig:flexce} for flexCE and Figure \ref{fig:lian} for the Lian model. As before, each column shows a different dwarf galaxy. The top row shows the best-fit model chemical track through the abundance pattern, the middle row shows the star formation history of that chemical track, and the bottom row shows the metallicity evolution as a function of time. We describe the results in the following subsections, and compare and contrast the model results in \S \ref{sec:chem_sum}. Because the models are only parameterized with a single burst in SF after the initial SF epoch, the models are likely fitting to the most significant ``burst'' in a galaxy's SFH that provided the most stars with the ages that we probe with APOGEE.

\subsubsection{Magellanic Clouds}

The flexCE results show that strong bursts are favored to match the abundance patterns of both the LMC and SMC, although the burst in the SMC is somewhat weaker, and occurred $\sim$ 4 Gyr earlier than the burst in the LMC. The burst results in quick metallicity evolution, with the LMC enriching from [Fe/H] $\simeq$ -0.8 to [Fe/H] $\simeq$ -0.3 in the last 2-3 Gyr. The SMC experienced similar swift enrichment, but over a period of 5-8 Gyr ago, with the evolution much slower in the last 2-3 Gyr. 

The Lian results also favor bursts in the MCs, although the bursts are more similar in relative strength between the two galaxies than the flexCE results. The Lian models also predict a ``dip'' in star formation that reached a minimum at 10 Gyr for the LMC and at 8 Gyr for the SMC, and show that the duration of the burst is shorter for the SMC than the LMC. Like the flexCE results, the Lian model results favor a scenario in which the SMC experienced a peak star burst $\sim$ 4 Gyr before the LMC. Also like the flexCE models, the Lian models show that the bursts result in rapid metallicity evolution, although the metallicity evolution is more rapid than the flexCE results, and begins at [Fe/H] = -1.3 rather than at [Fe/H] = -0.8. This is at least partially due to the difference in how the two models are parameterized.

The flexCE models predict that all LMC stars with [Fe/H] $\gtrsim$ -0.7 should be younger than 3-4 Gyr, and the Lian models predict the same age for stars with [Fe/H] $\gtrsim$ -1.2. The [C/N] results in \S \ref{sec:cn_age} suggest that the LMC stars at [Fe/H] $>$ -0.6 are significantly younger than the MW thin disk stars at the same metallicity, which is qualitatively consistent with these predictions.  

\subsubsection{GSE}
Both models find a SFH for GSE that peaks at early times and declines. Because the gas inflow is treated differently in the two models, the flexCE models show a broader SF peak located at a slightly later time, than the Lian model. While the model tracks extend to present day, GSE reaches [Fe/H] = -0.5 (maximum observed metallicity) $\sim$ 5 Gyr into its evolution in the flexCE models and 8-9 Gyr into its evolution in the Lian models, although the Lian models show very slow GSE metallicity enrichment beginning at  $\sim$5 Gyr into its evolution. This is consistent with the picture that GSE merged with the MW some 8-10 Gyr ago, thus cutting off this late evolution. Our results prefer a merger some 7-9 Gyr ago.

In the flexCE models we adopt similar inflow timescales for GSE and the MCs (2.5 Gyr vs. 2 Gyr), and the fit leads to a much higher (factor 10-20) SFE and thus to a SFH that peaks at much earlier times. This result is driven by the higher metallicity of the ``knee'' in the [Si/Fe]-[Fe/H] diagram. According to the flexCE SFH, by 5 Gyr into its evolution, GSE achieved [Fe/H] = -0.5, as compared to -1.2 and -1.5 for the LMC and SMC, respectively. The Lian models show a similar result, with GSE enriching to [Fe/H] = -0.8 5 Gyr into its evolution as compared to -1.4 and -1.55 for the LMC and SMC, respectively. Here the SFE is similar between the MCs and GSE, and the strong early SF peak in GSE arises from a short inflow timescale. In the flexCE model, it took the LMC 10 Gyr to enrich to [Fe/H] = -0.7, which is about where GSE enriched to before merging with the MW. In the Lian model, the LMC only attained GSE metallicities at recent times, although GSE took $\sim$6 Gyr to reach [Fe/H] = -0.7 instead of $\sim$4 Gyr in the flexCE model.

\subsubsection{Sgr}
As described in \S \ref{sec:chem_results}, the relatively flat $\alpha$-element abundance patterns at [Fe/H] $>$ -0.8 suggest some kind of extended SFH. Both models find that a second star formation epoch, beginning some 5-6 Gyr into its evolution and rising above the declining trend from earlier times, is required in Sgr to produce the flatter or ``hooked'' abundance $\alpha$-element abundance pattern. Without such a secondary peak, the chemical evolution tracks show monotonically decreasing [Si/Fe] or [Mg/Fe] with increasing [Fe/H], rather than the slight flattening of the observations. In the flexCE model, the sharp minimum of star formation at 5 Gyr produces the kink in the [Si/Fe]-[Fe/H] abundance track, slightly improving the fit to the data. In both models, this second SF epoch is not as strong relative to the earlier epoch, in contrast to the MCs, and as such, the [$\alpha$/Fe] ratio is not enhanced like it is in the MCs. 

The early chemical enrichment of Sgr was in between GSE and the MCs in both models (e.g., intermediate SFE), with Sgr enriching to [Fe/H] = -0.7 and -1.0 5 Gyr into its evolution for the flexCE model and the Lian model, respectively. After this point, the flexCE model favor a shorter, stronger burst, compared to a more sustained, weaker burst for the Lian model. In both models, Sgr effectively stops forming stars 10 Gyr into its evolution, which is consistent with observations of Sgr that show few or no young (age $<$ 2-3 Gyr) stars (e.g., \citealt{Siegel2007}), as well as the [C/N] inferences shown in \S \ref{sec:cn_age}.

\subsubsection{Fnx}

As was the case for Sgr, both the flexCE and Lian models find support for a second epoch of star formation in Fnx, although this second ``burst'' was much weaker than the initial SF peak ($\sim$ 3 times weaker for the flexCE results and $\sim$ 9 times weaker for the Lian model results). Even though Fnx has a much lower stellar mass than the other galaxies, both models suggest that Fnx enriched to nearly the same [Fe/H] as the SMC 5 Gyr into its evolution.

Like the MCs, Fnx is one of the galaxies where the [Mg/Fe]-[Fe/H] and [Si/Fe]-[Fe/H] patterns differ most strongly from each other, which is likely a source of at least some of the differences in results between the two chemical evolution codes.  

\subsection{Summary and Model Comparison}
\label{sec:chem_sum}

\begin{figure}
\includegraphics[width=1.0\hsize,angle=0]{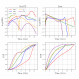}
\caption{Comparison of the derived SFHs for the two different model sets, flexCE on the left and Lian on the right. The top rows shows the SFR as a function of time, and the bottom rows shows the cumulative star formation over time. } 
\label{fig:model_compare}
\end{figure}

In Figure \ref{fig:model_compare} we show the SFHs of each galaxy for the flexCE results and Lian model results, comparing the SFR (top row) and the cumulative star formation (bottom row). The top row of Figure \ref{fig:model_compare} is useful to compare the relative shapes of the SFHs of the galaxies, but because the mass scale of the models has not been adjusted such that the final stellar mass of the model matches the observed stellar masses of each galaxy (see \S \ref{sec:flexce}), the normalization/scaling of the SFR of each galaxy is inaccurate. 

The final stellar mass of each system in the flexCE models, for example, do, however, match the mass ordering that observations suggest, with the final stellar masses of $7.3 \times 10^9$ M$_{\odot}$ for the LMC, $2.3 \times 10^9$ M$_{\odot}$ for the SMC, and $3.7 \times 10^8$ M$_{\odot}$ for Fnx.  The final stellar masses of the flexCE models for Sgr, $2.0 \times 10^9$ M$_{\odot}$, and GSE, $2.9 \times 10^9$ M$_{\odot}$ after 5 Gyr of evolution, are also consistent in suggesting that Sgr and GSE may have been similar in mass to the SMC prior to their respective disruption (see the discussion in \S \ref{sec:mass_environment}).  Nonetheless we remark that the resulting final stellar masses are $\sim 5-10\times$ too high compared to observational estimates (see \S \ref{sec:mass_environment} for observed masses; but note again as mentioned in \S \ref{sec:flexce}, the total mass scale does not impact the final chemical tracks).  Therefore, a better way to compare SFHs across galaxies from the models is to use the cumulative star formation histories (CSFH), shown in the bottom row of Figure \ref{fig:model_compare}.

Both sets of models show weaker early SFH for the MCs as compared to the other galaxies, and both models show that the MCs experienced peaks in SF that occurred much later than the SF peaks in the other galaxies, with the most significant period of enhanced SF in the SMC occurring $\sim$ 4 Gyr before the LMC. 

The model SFHs differ in the relative strengths of the bursts, with the SMC, Sgr, and Fnx all showing weaker bursts in the Lian results than the flexCE results. The Lian models also imply that GSE and Sgr each formed about 70\% of their stars in the first 4 Gyr of evolution, whereas the flexCE models show Sgr forming 60\% of its stars prior to this point, and GSE 40\%. However, both models predict GSE evolution continuing beyond the observed tracks, and stopping them at the maximum observed [Fe/H] $\simeq$ -0.5 implies truncating star formation at $t\simeq$ 5 Gyr in the flexCE model and $t\simeq$ 10 Gyr in the Lian model, a plausible result of disruption or ram pressure stripping by the MW. The substantially weaker burst for Fnx in the Lian models also results in nearly all stars in Fnx forming by 6 Gyr into its evolution, whereas the flexCE models find only 70\% of the stars to have formed at this point.

\section{Discussion}
\label{sec:disc}

In the following section we discuss what these SFHs mean in the greater context of galaxy formation and evolution. We also compare our SFH results from chemistry alone to those derived in the literature primarily through photometry, as well as discuss future prospects. 

\subsection{Mass and Environmental Effects on Galaxy Evolution}
\label{sec:mass_environment}
The chemical abundance patterns and inferred SFHs highlight the importance of galaxy environment for chemical evolution. \citet{McConnachie2012} and references therein tabulate the following stellar masses for the galaxies in our sample: LMC = $1.5 \times 10^{9} M_{\odot}$, SMC = $4.6 \times 10^{8} M_{\odot}$, Sgr (main body) = $2.1 \times 10^{7} M_{\odot}$, and Fnx = $2.0 \times 10^{7} M_{\odot}$. Sgr was much more massive in the past, with some studies finding  masses as high as $6.4 \times 10^{8} M_{\odot}$ (e.g., \citealt{Niederste-Ostholt2010}). Estimates for the stellar mass of GSE are generally in the range $3-10 \times 10^{8} M_{\odot}$ (e.g., \citealt{Mackereth2019a}).

The LMC, GSE, SMC, and Sgr span about an order of magnitude in stellar mass, with the LMC being the most massive. Fnx is at least another order of magnitude less massive even than the SMC/GSE/Sgr. Despite its relatively low present-day mass compared to the other galaxies, Sgr is the most enriched dwarf galaxy, reaching slightly higher metallicities than the LMC. The SMC, GSE, and Fnx have all enriched to nearly the same metallicity, despite their vastly different stellar masses. What is very different across these dwarfs is the environments in which they formed and evolved.

Currently, these galaxies are all well inside the MW environment, or have already been accreted by the MW. However, GSE is thought to have formed close to the MW and was accreted at early times (e.g., \citealt{Gallart2019,Mackereth2019a}), Sgr and Fnx at more intermediate times (e.g., \citealt{Rocha2012,Fillingham2019}), and the MCs are only falling in recently, having likely evolved in near isolation until now (e.g., \citealt{Besla2016}). In the paradigm of \citet{Gallart2015}, in which dwarfs galaxies are assigned to two groups according to their SFH, we could consider the MCs to be ``slow dwarfs'', forming the bulk of their stars more recently, as compared to the ``fast dwarfs'' (e.g., Sgr, GSE, and Fnx), that started their evolution with a dominant early star formation event  (see e.g., \citealt{Shi2020} for similar simulation results). However, even the MCs became ``faster'' dwarfs when they began to interact with each other driving up the $\alpha$-element abundances. This suggests that proximity to any galaxy, not just a more massive Milky Way, can be important drivers in star formation history. Had the MCs not interacted with each other, they would likely have a lot of gas, still, but contain far fewer stars. The inferred SFH presented here suggests that without the interactions between them, the MCs would have enriched to [Fe/H] $\simeq$ -0.7 for the LMC and [Fe/H] $\simeq$ - 1.5 for the SMC.

There is good agreement in the literature that the MW and M31 environments have strong effects on the quenching times of in-falling satellite galaxies, with many low-mass galaxies (M$_{*}$ $<$ $10^{8} M_{\odot}$) quenching within $\sim$2 Gyr of passing through the virial radii of the host galaxies (e.g., \citealt{Rocha2012,Slater&Bell2014,Fillingham2015,Weisz2015,Wetzel2015,Akins2021}). This is likely a consequence of ram pressure stripping effectively removing gas from these galaxies, preventing further star formation (see e.g., \citealt{Hester2006,Bekki2014,Fillingham2016}). Sgr was likely just massive enough to avoid fast quenching from ram pressure stripping (e.g., \citealt{Niederste-Ostholt2010}), but the mass of Fnx is low enough (M$_{*}$ $\sim 10^{7} M_{\odot}$, e.g., \citealt{McConnachie2012}) such that it should have been quickly quenched when entering the environment of the MW. However, the extended SFH of Fnx inferred from both the APOGEE data presented here and other literature studies can be reconciled with its low mass if the environmental quenching timescale also depends on satellite orbit. Recent studies have found that not all of these low-mass dwarfs are quenched, with the galaxies on more circular orbits and larger pericenter passages (such as Fnx) having much longer quenching timescales (e.g., \citealt{Fillingham2019}). Such ``fortuitous'' orbits have been found to enhance star formation in some simulations (e.g., \citealt{DiCintio2021}), and observational evidence exists of Sgr and the MW undergoing enhanced SF epochs, coincident with Sgr pericenter passages \citep{Ruiz-Lara2020}).

In addition to environment, the mass of a galaxy can also play an important role in its evolution. The less massive dwarf spheroidal galaxies, such as Sculptor, Draco, and Ursa Minor, have formed very few stars in recent times, and have only enriched to [Fe/H] $\sim$ -1.0 or lower. Moreover, there is an established mass-\emph{mean} metallicity relationship in the literature described in \citet{Kirby2011}, with more massive galaxies enriching to a higher mean metallicity, plausibly because they are able to retain some of their gas that is ejected from SNe. Because of APOGEE selection biases, it is difficult to use the APOGEE mean metallicities to accurately place these galaxies on the mass-metallicity relationship. However, Sgr likely has a much higher metallicity for its present-day mass as compared to the SMC, and maybe even as compared to the LMC. Fornax also likely has a mean metallicity that is closer to the SMC than perhaps it should be for its mass. Conversely, another way to view the discrepancy is that the MCs, having evolved in relative isolation until recent times, are too metal-poor for their large stellar masses. While works such as \citet{Reichert2020} and \citet{Hendricks2014} have shown that Local Group galaxies tend to exhibit correlations between their $\alpha$-element enrichment and luminosity, the MCs would also be an exception to this correlation (see also \citet{Nidever2020}).

\citet{Geha2012} find that nearly all field galaxies with a stellar mass $<$ $10^{9} M_{\odot}$ are still forming stars today. Perhaps mass is the primary fundamental driver for how enriched a given galaxy is, but galaxy-galaxy interactions can push galaxies off of this relation, either by merging and increasing the mass without too much extra star formation, or by falling into the potential of a massive galaxy, where star formation is momentarily kick started before the galaxy is disrupted or quenched.

\subsection{Comparison to Photometric Star Formation Histories}
\label{sec:photo_comp}

This work is far from the first to derive star formation histories for these galaxies. However, this is among the first work to derive/estimate star formation histories from the chemical abundance patterns of multiple galaxies, applying the same model frameworks to observations from nearly identical setups. In the following section we compare the star formation histories we measure above to those measured in the literature. Throughout this section we use ``photometric star formation histories (SFHs)'' to refer to star formation histories derived from color-magnitude diagrams. As described in greater detail below, these photometric SFHs can vary in precision depending on whether or not the photometry is deep enough to reach the old main sequence turn-off (oMST, e.g., \citealt{Ruiz-Lara2018}). Moreover, most of the literature photometric studies for these galaxies are small, pencil-beam patches of the galaxies (typically the central regions) as compared to our samples, which cover large fractions of the entire spatial extent of these galaxies.

\subsubsection{The Magellanic Clouds}

\begin{figure}
\includegraphics[width=1.0\hsize,angle=0]{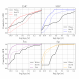}
\caption{Comparison between the SFHs derived in this work and the SFHs derived in \citet{Weisz2013} (LMC and SMC, top row) and \citet{Weisz2014} (Fnx and Sgr, bottom row).  }
\label{fig:weisz_comp}
\end{figure}

\begin{figure}
\includegraphics[width=1.0\hsize,angle=0]{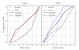}
\caption{Comparison of our inferred cumulative SFHs to those of \citet{Harris&Zaritsky2009}, \citet{Harris&Zaritsky2004}, and SMASH \citep{Nidever2021} for the LMC (left) and SMC (right).}
\label{fig:mc_hz_compare}
\end{figure}

There have been many photometric studies of the SFHs of the MCs. In this section we compare the MC chemical SFHs to the photometric studies of Weisz (\citealt{Weisz2013,Weisz2014}, Figure \ref{fig:weisz_comp}), Harris and Zartitsky (HZ, \citealt{Harris&Zaritsky2004,Harris&Zaritsky2009}, Figure \ref{fig:mc_hz_compare}), and  more recent SFHs derived from the Survey of the MAgellanic Stellar History (SMASH, \citealt{Nidever2017,Nidever2021}, Figure \ref{fig:mc_hz_compare}). The HZ work is ground-based, so is typically shallower photometry but covers much of the same spatial regions of the MCs as our data. The Weisz work is much deeper photometry from HST, but is of the central regions of the MCs. The SMASH work is both deep and covers large spatial regions of the MCs.

We find reasonable agreement between the various photometric SFHs of the LMC and what we infer from the chemical evolution models fit to the APOGEE data. In the upper-left panel of Figure \ref{fig:weisz_comp}, we find that the Weisz cumulative SFH falls between what is derived for the two models from this work for the LMC at older ages, but all 3 results are in agreement with the LMC forming $\sim$20\% of its stars in the last 2 Gyr. This is similar to the SFHs from HZ and SMASH (shown in the left panel of Figure \ref{fig:weisz_comp}), although the SMASH SFHs suggest that the LMC formed $\sim$40\% of its stars by 4 Gyr into its evolution, which is slightly higher than Weisz, and much higher than the $\sim$10\% found in the HZ work. This discrepancy is possibly due to the fact that the HZ work was not deep enough to capture the oMST, resulting in more uncertain SFRs at early times. The Lian model agrees well with the SMASH LMC SFH, and the flexCE agrees well with the \citet{Harris&Zaritsky2009} LMC SFH. 

The SFH results for the LMC derived by \citet{Monteagudo2018} are qualitatively similar to what is shown here. They find some spatial dependence of the SFH, with the bar region of the LMC forming a higher fraction of total stars at more recent times than the disk. Specifically, they find that the disk regions of the LMC formed half of their stars 7-8 Gyr into its evolution, whereas the bar had not formed half of their stars until 9-10 Gyr into the galaxy's evolution. \citet{Meschin2014} find similar results, and also that the trend of larger fraction of stars formed at earlier times extends to the outer regions of the LMC. A detailed comparison of the spatial dependence of the SFH is beyond the scope of this work.

For the SMC, we find worse agreement in our inferred SFHs compared to the Weisz SFH than we do for the LMC (upper-right panel of Figure \ref{fig:weisz_comp}). Specifically, both models find a much larger fraction of SMC stars forming at earlier times than the \citet{Weisz2013} work. This could be in part due to the difference in spatial coverage, which is why there is slightly better agreement with the SMASH work shown in the right panel of Figure \ref{fig:mc_hz_compare}, where the flexCE, Lian, and SMASH CSFHs all show that the SMC formed $\sim$70-80\% of its stars in the first 8 Gyr of its evolution. A recent study by \citet{Rubele2018} suggests significant enhancement of SF took place in the SMC beginning 100 Myr ago. Our sample selection cuts nearly all of these stars out (see \S \ref{app:targ_mc}), but these stars still represent a small fraction of total stars formed in the SMC. Future analysis confirming the reliability of results for massive supergiants in APOGEE will allow for a chemical exploration of these more massive, younger stars.

\subsubsection{Sgr and Fnx}

We compare the Weisz Sgr and Fnx SFHs to those that we infer in the bottom row or Figure \ref{fig:weisz_comp}. The flexCE Fnx model reproduces a cumulative star formation history that is very much like what is derived in \citet{Weisz2014}, with Fnx forming 20\% of its stars in the last $\sim$ 4 Gyr. The Lian model, with the much weaker secondary SF epoch, implies Fnx formed the vast majority of its stars by 6 Gyr ago. Support for a recent burst in Fnx can be found throughout the literature, including the recent work of \citet{Rusakov2021}, who find that Fnx underwent a strong starburst some 4-5 Gyr ago, one that was close to or even exceeded the strength of the early SF event. This is more consistent with our flexCE results, which shows a nearly equal strength second burst in Fnx that peaked 4-5 Gyr ago. \citet{Hendricks2014} find chemical evidence for a starburst in Fnx, like we do here in this work, but they suggest the burst occurred at much earlier times than either the flexCE or Lian models suggest. The more spatially-extended SFH work of \citet{deboer2012} shows Fnx underwent a more gradual enrichment from 3-8 Gyr ago, enriching from [Fe/H] = -1.5 to [Fe/H] = -0.5, which is an enrichment rate that is more consistent with the flexCE results.

The Sgr comparison is shown in the bottom-right panel of Figure \ref{fig:weisz_comp}. Neither model matches well with the \citet{Weisz2014} SFH of Sgr, though they agree well with each other. This could be due to the fact that our Sgr sample contains many stream stars, whereas \citet{Weisz2014} was looking at the main body of the galaxy, where the tidal interactions have preferentially removed many metal-poor stars, including into the streams. However, all of the star formation histories agree that Sgr formed nearly all of its stars by 3-4 Gyr ago. This is mostly in line with other photometric SFH studies, such as \citet{Siegel2007}, although we do not find the metal-rich youngest populations that they find in the core. \citet{deBoer2015} find that the Sgr stream stars exhibit a tight age-metallicity relation, enriching to [Fe/H] = -0.7 by 5-7 Gyr ago, again consistent with both model results (see the bottom row of Figures \ref{fig:flexce} and \ref{fig:lian}). Our SFH here also qualitatively agrees with recent work by \citet{Garro2021}, who analyzed the ages and metallicities of the globular cluster population of Sgr, including 12 new clusters discovered by \citet{Minniti2021}. They find that Sgr formed its metal-rich (-0.9 $<$ [Fe/H] $<$ -0.3) globular clusters some 6-8 Gyr ago.

\subsection{Future Prospects}
\label{sec:future}

We have restricted the parameter space of our chemical evolution models in part for computational practicality in this exploratory study, but also because we are constraining them with a single observable, the [Si/Fe]-[Fe/H] or [Mg/Fe]-[Fe/H] track. In future work, we can allow greater model flexibility by simultaneously employing these observables (including more than just two chemical elements) and the photometric star formation histories illustrated in Figures \ref{fig:weisz_comp} and \ref{fig:mc_hz_compare}, and we can test or further constrain these models using metallicity distribution functions (MDFs). Computing MDFs from the metallicity distribution of observed APOGEE stars requires correcting for selection biases; alternatively, one can incorporate selection effects into the model and directly predict the observed distributions. For completeness, we show and discuss the uncorrected MDFs in \S \ref{app:mdfs}.

One natural route for such an investigation is to take the photometrically-inferred star formation histories as a starting point, then derive the gas accretion history that produces this star formation for given assumptions about star formation efficiency. Fitting the evolutionary tracks will restrict the model parameters, and model variations that predict different enrichment vs. time will predict different MDFs. As our exploratory results already suggest, chemical evolution constraints may make it possible to detect bursts or other variations in the star formation history. Joint modeling can test the need for more radical differences among the galaxies being considered, such as different IMFs, different Type Ia SNe populations, different AGB yields that could arise from systematic differences in stellar rotation, or differences in outflow physics such as preferential ejection of Type II SNe products relative to Type Ia SNe or AGB products.

\section{Conclusions}
\label{sec:conc}

We have presented the APOGEE chemical abundance patterns for five MW dwarf galaxies that span a range in mass and evolution environments. Our major conclusions are summarized as follows:

\begin{itemize}
    \item The chemical abundance patterns of these five dwarf galaxies show very different early star formation efficiencies,  with GSE having had the strongest, followed by Sgr, then the MCs. Fnx plausibly had an early star formation efficiency similar to Sgr, but exhibits the lowest $\alpha$-element abundance at [Fe/H] = -1.2, suggesting either a rapidly declining SFH or outflows that preferentially ejected Type II SNe products.
    \item All dwarf galaxies except for GSE show chemical signs of extended star formation periods after the initial star formation epochs, with the LMC showing an increasing [$\alpha$/Fe]-[Fe/H] abundance pattern, suggestive of a more intense starburst compared to the other galaxies. The lower [C/N] abundances of the metal-rich LMC stars suggest that they were formed at more recent times than the Sgr and MW low-$\alpha$ sequence stars at the same metallicity.
    \item In median abundance trends, [X/Fe]-[Fe/H] and [X/Mg]-[Mg/H], Fnx is the strongest outlier among these five dwarfs, followed by GSE. Tracks for the MCs and Sgr are fairly similar except for low-metallicity [$\alpha$/Fe] differences driven by star formation efficiency, and high-metallicity [$\alpha$/Fe] differences driven by late star formation.
    \item The C+N and Ce abundance patterns show that all galaxies had greater contribution of AGB enrichment (relative to SNe contributions) to their evolution as compared to the MW high-$\alpha$ sequence, with Fnx showing the highest relative contributions and GSE showing the lowest at [Mg/H] $<$ -0.8. However, the most metal-rich Sgr stars have the highest [Ce/Mg] abundances of any galaxy, and therefore have had the largest relative contribution of AGB stars to their enrichment.
    \item The deficient [Al/Mg] abundances of the dwarf galaxies relative to the MW is plausibly a result of the overall lower metallicity Type II SNe that occurred in the dwarf galaxies as compared to the MW. Lower star formation efficiency may lead to a great ``lag'' between the metallicity of the ISM and the metallicity of the stars that enriched the ISM.
    \item The deficient [Ni/Fe] abundances of the dwarf galaxies relative to both MW sequences combined with the deficient [Ni/Mg] relative to the MW low-$\alpha$ sequence can be explained if the production of Ni is dependent on the metallicity of the supernova progenitor.
\end{itemize}

We also use chemical evolution models to quantify some aspects of these chemical abundance patterns, finding that the SMC also experienced a recent enhancement of star formation, but this enhancement was weaker than the burst in the LMC by a factor of $\sim$2-3, and occurring 2-4 Gyr before the burst in the LMC. While not found to be coincident in the chemical evolution models, it is likely that the increased star formation epochs in both galaxies are results of their mutual interactions.

From the models, we infer similar extended star formation events occurring in Sgr and Fnx some 3-6 Gyr ago, but these increases in star formation are small compared to the initial bursts. Future works that are able to properly account for selection biases can combine the photometric constraints on the SFHs with the chemical abundance patterns and metallicity distribution functions to investigate further details of the SFHs of these galaxies (e.g., IMF variation, different AGB yields, etc.).

This comparative chemical abundance analysis highlights the role galactic environment has on shaping a galaxy's chemical evolution. The most isolated galaxies, the MCs, had the weakest early star formation efficiency whereas Sgr and GSE enriched to much higher metallicities before Type Ia SNe began to significantly contribute to their chemical enrichment histories, likely due to their proximity to the MW. The less-massive Sgr was able to continue forming stars upon beginning its merger with the MW whereas the evolution of GSE was likely cut short as it merged, perhaps because of its larger mass and more radial infall trajectory. The MCs evolved slowly in isolation, before interacting with each other at more recent times to drive up their star formation. Fnx was apparently not quenched when it fell into the MW environment, showing chemical signatures of a star formation history that was extended by either a merger, or a pericenter passage around the MW.

\acknowledgments
SH is supported by an NSF Astronomy and Astrophysics Postdoctoral Fellowship under award AST-1801940. JL is supported by the National Science Foundation under Grant No. 2009993. DW acknowledges support of NSF grant AST-1909841. MRC acknowledges funding from the European Research Council (ERC) under the European Union’s Horizon 2020 research and innovation programme (grant agreement no. 682115). DM is supported by the BASAL Center for Astrophysics and Associated Technologies (CATA) through grant AFB-170002, and by FONDECYT Regular grant No. 1170121. CG acknowledges support from the State Research Agency (AEI) of the Spanish Ministry of Science, Innovation and Universities (MCIU) and the European Regional Development Fund (FEDER) under grant AYA2017-89076-P. DAGH acknowledges support from the State Research Agency (AEI) of the Spanish Ministry of Science, Innovation and Universities (MCIU) and the European Regional Development Fund (FEDER) under grant AYA2017-88254-P. R.R.M. acknowledges partial support from project BASAL AFB-$170002$. D.G. gratefully acknowledges support from the Chilean Centro de Excelencia en Astrof\'isica y Tecnolog\'ias Afines (CATA) BASAL grant AFB-170002. D.G. also acknowledges financial support from the Direcci\'on de Investigaci\'on y Desarrollo de la Universidad de La Serena through the Programa de Incentivo a la Investigaci\'on de Acad\'emicos (PIA-DIDULS). MR acknowledges funding from UNAM-DGAPA PAPIIT IN109919, CONACyT CF-2019-86367, and CY-253085. T.C.B. acknowledges partial support for this work from grant PHY 14-30152:  Physics Frontier Center / JINA Center for the Evolution of the Elements (JINA-CEE), awarded by the US National Science Foundation.

Funding for the Sloan Digital Sky Survey IV has been provided by the Alfred P. Sloan Foundation, the U.S. Department of Energy Office of Science, and the Participating Institutions. SDSS-IV acknowledges support and resources from the Center for High-Performance Computing at the University of Utah. The SDSS web site is www.sdss.org.

SDSS-IV is managed by the Astrophysical Research Consortium for the Participating Institutions of the SDSS Collaboration including the Brazilian Participation Group, the Carnegie Institution for Science, Carnegie Mellon University, the Chilean Participation Group, the French Participation Group, Harvard-Smithsonian Center for Astrophysics, Instituto de Astrof\'isica de Canarias, The Johns Hopkins University, Kavli Institute for the Physics and Mathematics of the Universe (IPMU) / University of Tokyo, Lawrence Berkeley National Laboratory, Leibniz Institut f\"ur Astrophysik Potsdam (AIP), Max-Planck-Institut f\"ur Astronomie (MPIA Heidelberg), Max-Planck-Institut f\"ur Astrophysik (MPA Garching), Max-Planck-Institut f\"ur Extraterrestrische Physik (MPE), National Astronomical Observatories of China, New Mexico State University, New York University, University of Notre Dame, Observat\'ario Nacional / MCTI, The Ohio State University, Pennsylvania State University, Shanghai Astronomical Observatory, United Kingdom Participation Group, Universidad Nacional Aut\'onoma de M\'exico, University of Arizona, University of Colorado Boulder, University of Oxford, University of Portsmouth, University of Utah, University of Virginia, University of Washington, University of Wisconsin, 
Vanderbilt University, and Yale University.

This research made use of Astropy \footnote{http://www.astropy.org} a community-developed core Python package for Astronomy \citep{astropy:2013,astropy:2018}, SciPy \citep*{SciPy}, NumPy \citep{NumPy}, and Matplotlib \citep{Hunter:2007}.

\bibliographystyle{apj}
\bibliography{ref_og.bib}

\begin{appendix}

\section{Target Selection Supplemental}
\label{app:targ}

\subsection{Magellanic Clouds}
\label{app:targ_mc}

\begin{figure}[t]
\includegraphics[width=0.5\hsize,angle=0]{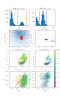}
\caption{Top row: radial velocity distribution for stars that fall within $12^{\circ}$ of the LMC center (left) and within $8^{\circ}$ of the SMC center (right). Red and blue vertical lines mark the RVs that are $\pm3\sigma$ from the median RV of the LMC and SMC, respectively, used as cuts to select potential members. Second Row: The PM distributions for the LMC (left) and SMC (right) colored by whether or not the stars fall within the RV cuts indicated in the top row. The red and blue boxes indicate the PM cuts applied to select the final MC members. Third Row: CMD for the stars that pass the RV and proper motion cuts. Stars are colored according to their APOGEE metallicity, as indicated by the color bar at the right. Red and blue lines indicate the photometric cuts used to remove obvious massive stars above the lines. Fourth Row: Same as the third row, but zoomed in to illustrate where the RSGs are as bright as the tip of the RGB. }
\label{fig:mc_targ}
\end{figure}

The Magellanic Clouds have been targeted through multiple programs in APOGEE, some targeting the young, massive stars in the clouds, and others sampling the giant branches (see \citealt{Zasowski2017}, \citealt{Nidever2020}, and \citealt{Santana2021} for all details). In this work we focus on only the red giant branch stars for which we know APOGEE is able to derive reliable abundances (\citealt{Jonsson2020}). To select our MC sample, we first make spatial cuts, selecting all stars with a projected spherical distance within 12$^{\circ}$ and 8$^{\circ}$ of the centers of the LMC and SMC, respectively. The centers we adopt are (80.893860$^{\circ}$,-69.756126$^{\circ}$) for the LMC and  (13.18667$^{\circ}$,-72.8286$^{\circ}$) for the SMC ($\alpha$, $\delta$). To remove obvious MW foreground contamination, we remove stars that are $\pm$ 3$\sigma$ from the median APOGEE-measured RV of each galaxy, as shown in the top row of Figure \ref{fig:mc_targ}. We then make similar $\pm$ 3$\sigma$ cuts in each proper motion dimension from \emph{Gaia} EDR3 to further remove MW contamination (second row of Figure \ref{fig:mc_targ}). 

As shown in the third row of Figure \ref{fig:mc_targ}, these cuts result in a mixture of upper RGB stars, AGB stars, luminous AGB-O stars, red supergiant (RSG) stars, massive blue main sequence stars,and even some objects around the instability strip. Because many of these types of objects are stars for which we do not know if the APOGEE abundance pipeline produces reliable results, we employ further cuts to select a sample of largely RGB stars. To do this, we first select stars below the tip of the red giant branch, as measured and defined by \citet{Hoyt2018}: $$H < (18.49-5.94-1.62\times[(J-K_{s})-1.00]).$$

We make the cut 0.1 mag brighter to account for varying depth of field across the galaxy. For the SMC, we use the same functional form, but account for the 0.5 difference in distance modulus. These selections are illustrated in the bottom four panels of Figure \ref{fig:mc_targ} as the red and blue lines for the LMC and SMC, respectively. We also exclude stars from both galaxies with ($J-K_{s}$) $>$ 1.3, to avoid obvious carbon stars.

While these photometric cuts remove most of the massive evolved stars (M $\gtrsim 3 M_{\odot}$), the bottom-left panel of Figure \ref{fig:mc_targ} shows that some of the faintest RSGs in the LMC still make it into the photometric selection. We remove those by requiring that all stars with ($J-K_{s}$) $<$ 1.0 and H $<$ 12.8 have [Fe/H] $<$ -0.55. Note that this photometric selection means that our sample is biased against the youngest stars (Age $\lesssim$ 1 Gyr).  

\subsection{GSE}
\label{app:targ_gse}

\begin{figure*}[t]
\includegraphics[width=1.0\hsize,angle=0]{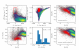}
\caption{Plots that highlight the dynamical and chemical selections used to select GSE stars. Red points indicate GSE candidates and the `viridis' density map shows the MW parent sample from which they are selected. Upper-Left Panel:  L$_{z}$-$\sqrt{J_{R}}$ plane where the initial dynamical selection is made. Upper-middle Panel: Energy-$L{z}$ plane to highlight where these candidates lie. Upper-Right Panel: [Mg/Fe]-[Fe/H] abundance distribution of the candidates. Bottom Row: additional chemical cuts applied in [(C+N)/Fe]-[Fe/H] space to remove MW high-$\alpha$ disk contamination. }
\label{fig:gse_targ}
\end{figure*}

To select the GSE sample, we start with the initial quality cuts described in \S \ref{sec:samp} and remove stars belonging to known globular clusters, also avoiding regions of the sky containing the Magellanic Clouds. Specifically, we do not include stars that have a projected distance of 12 degrees from the LMC and 8 degrees from the SMC. Then, considering only stars with [Fe/H] $<$ 0.0, we make kinematic cuts using the orbital angular momenta (L$_{z}$) and the square root of radial orbital action ($\sqrt{J_{R}}$), adopting the orbital properties computed with astroNN \citep{Leung2019}. We follow the work of \citet{Feuillet2020} and select stars with $|L_{z}| < 500$~km~kpc~s$^{-1}$ and $\sqrt{J_{R}} = 30-50$~(kpc~km~s$^{-1}$)$^{1/2}$, as shown by the red selection box in the upper-left panel of Figure \ref{fig:gse_targ}. The upper-middle panel of Figure \ref{fig:gse_targ} shows where these stars lie in the Energy-L$_{z}$ plane, which many other studies use to select GSE stars (e.g., \citealt{Myeong2018,Horta2021c,Naidu2021}). 

While this sample largely follows the expected [Mg/Fe]-[Fe/H] abundance pattern of GSE (e.g., \citealt{Hayes2018a,Haywood2018}) that is shown in the top-right panel of Figure \ref{fig:gse_targ}, there is clear contamination by the MW high-$\alpha$ ``thick disk'' stars. Therefore, we apply an additional [(C+N)/Fe] cut for stars with [Fe/H] $>$ -1.05, as demonstrated in the bottom row of Figure \ref{fig:gse_targ} and motivated by \citet{Hayes2018a}.

We find no obvious metallicity trend of our GSE members with L$_{z}$, suggesting we are not heavily biased in our sample. However, as mentioned in the text, should there be an undiscovered remnant of GSE that we are not observing here, then our comparison of GSE to the other galaxies is not complete.

\subsection{Sgr}
\label{app:targ_sgr}

To select Sgr members, we follow a method similar to that described in \citet{Hayes2020}. This work exploited the fact that the Sgr orbital plane is nearly perpendicular to the MW disk, making it easy to identify stars belonging to the Sgr core and stream. As demonstrated in \citet{Hasselquist2019b} and \citet{Hayes2020}, the APOGEE survey has observed hundreds of Sgr stream stars strewn across much of the sky.

We first transform the APOGEE sample into the Sgr coordinate system described in \citet{Majewski2003}. We then make initial cuts of:
\begin{itemize}
    \item $|\beta_{s}| < 30^{\circ}$ to remove stars out of the Sgr orbital plane.
    \item $d_{helio} > 10$ kpc to remove stars that are too close to be Sgr stream members.
    \item $[$Fe/H$]$ $<$ 0.0 to remove distant MW stars in and behind the bulge that are too metal-rich to be Sgr stream members.
\end{itemize}

\begin{figure*}[t]
\includegraphics[width=0.9\hsize,angle=0]{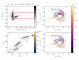}
\caption{Top-left Panel: $V_{zs}$-$L_{zs}$ plane showing the initial selection box, similar to what was done for \citet{Hayes2020}. Top-right panel: Galactic $Z_{\rm GC}$-$X_{\rm GC}$ plane of all stars in the top-left panel, with the selected stars colored by $\phi_{\rm vel, s}$, the velocity direction in the $V_{\rm xs}$ and $V_{\rm ys}$ velocities. Bottom-left panel: $\phi_{\rm vel, s}$ plotted as a function of longitude along the Sgr stream ($\Lambda_{s}$). The red ``A'', ``B'', and ``C'' regions indicate additional cuts placed on the sample as described in the text. Points are colored by heliocentric distance. Bottom-right panel: same as the top-right panel with the additional contamination removed. }
\label{fig:sgr_targ}
\end{figure*}

From these cuts, we then analyze the resulting distribution in the $V_{\rm zs}$ - $L_{\rm zs}$ plane, which is the velocity in the Z direction of the Sgr coordinate system plotted against the angular momenta in the Sgr system, shown in the upper-left panel of Figure \ref{fig:sgr_targ}. In principle, Sgr stars should have a $V_{\rm zs}$ velocity distribution centered around zero, and angular momenta consistent with the galactocentric distance of Sgr multiplied by its orbital velocity (i.e., 18 kpc x 270 km/s $\simeq$ 5000 kpc km/s). In practice, the distance uncertainties result in a structure where the two quantities are correlated. Still, we use the density map to select stars with $L_{\rm zs} > 1500$ kpc km/s, and -120 km/s $< V_{\rm zs} <$ 220 km/s. 

These cuts result in a spatially coherent core and trailing/leading arm structures, shown in the upper-right panel of Figure \ref{fig:sgr_targ}. The points are colored by $\phi_{\rm vel,s}$, the direction of the velocity vector in the Sgr X and Y coordinates. Stars that are colored the same are stars that are moving in the same direction. From this plot, we see some stars at Z$_{\rm{GC}} \sim$ 18 kpc and -10 kpc $<$ X$_{\rm{GC}}$ $<$ 10 kpc that are moving perpendicular to the stream stars found at slighly larger distances. These are likely halo contamination. 

We remove these stars by looking more closely at the $\phi_{\rm vel,s}$ distribution as a function of Sgr longitude ($\Lambda_{s}$), as was also done in \citet{Hayes2020}. In the lower-left panel of Figure \ref{fig:sgr_targ} we define three regions (A, B, and C) where there is likely contamination, removed according to the following prescriptions:

\begin{itemize}
    \item A: Stars that are moving perpendicular to the expected stream at these latitudes, and in direction of the bulge, therefore likely to be MW contamination.
    \item B: Stars that fall below the B line and have distance $<$ 30 kpc are likely not stream stars, as they are roughly the same distance as the stream, but moving perpendicularly. However, we do include the small handful of stars that fall below this line, but are at $d_{\rm{helio}} > 60 $ kpc, as these could be more distant Sgr stream structures.
    \item C: Stars that fall below the dashed C line, but are at $d_{\rm{helio}} > 50 $ kpc. 
\end{itemize}

\begin{figure}[t]
\includegraphics[width=0.6\hsize,angle=0]{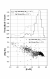}
\caption{Top: Metallicity distribution functions (MDFs) of the Sgr ``main body'' sample (solid black line) and the Sgr ``stream'' sample (dashed grey line). Bottom: [Mg/Fe]-[Fe/H] abundance plane with the Sgr main body sample plotted as black circles and the Sgr stream sample plotted as grey ``x''s.}
\label{fig:sgr_core_vs_stream}
\end{figure}

Note that there are $\sim$ 50 stars removed in total this way across the three regions, which constitutes only 5\% of the sample. We have confirmed that the inclusion or removal of these stars do not change our results. The lower-right  panel of Figure \ref{fig:sgr_targ} shows the final spatial distribution of our Sgr sample. While our Sgr sample consists of stars across much of the sky, $\sim$ 2/3 of our Sgr sample comes from the Sgr ``main body'' region, defined here as stars that are at a projected distance less than 12$^{\circ}$ from the center of Sgr. This main body region is shown in the inset figure of Figure \ref{fig:map}. While we do find that the MDFs of the main body and stream regions differ substantially (see also \citet{Hayes2020}), we show in bottom panel of Figure \ref{fig:sgr_core_vs_stream} that the [Mg/Fe]-[Fe/H] abundance tracks for the two regions do not differ significantly where they overlap in [Fe/H].

\subsection{Fnx}
\label{app:targ_fnx}

\begin{figure*}[t]
\includegraphics[width=0.4\hsize,angle=0]{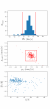}
\caption{Top panel: radial velocity distribution of the APOGEE ``FORNAX'' field. The red vertical lines indicate $\pm$3$\sigma$ from the median radial velocity of the field. Middle panel: \emph{Gaia} EDR3 proper motions of the same stars. A selection box of $\pm$3$\sigma$ in each direction of proper motion is used to remove potential contamination. Bottom panel: [Fe/H] vs. S/N for the sample. }
\label{fig:fnx_targ}
\end{figure*}

The Fornax selection is shown in Figure \ref{fig:fnx_targ}. APOGEE's Fornax field was specifically designed to target as many known members (based on previous radial velcocity studies and \emph{Gaia} proper motions) as possible, along with additional targets that were likely members by photometry only. We therefore clean the sample in a similar manner to the MCs (\S \ref{sec:targ_mc}). First, we only include stars that belong to the ``FORNAX'' APOGEE field. We then remove stars $>$ $\pm$ 3$\sigma$ from the median APOGEE RV, and then make a second selection on the \emph{Gaia} EDR3 proper motions (0.17 $< \mu_{\alpha} < 0.60$ and $-0.71 < \mu_{\delta} < -0.05$), as shown in the top two panels of Figure \ref{fig:fnx_targ}. These RV and PM cuts only remove some 12 stars from the Fornax plate.

Because Fornax is a distant galaxy, the S/N of these stars are generally much lower than those in the other galaxies. This means that the Fnx chemical abundances are more uncertain, as indicated by the error bars in \S \ref{sec:results} and Figures \ref{fig:mw_comp} and \ref{fig:mw_comp_mg}. The lower panel of Figure \ref{fig:fnx_targ} shows that our adopted lower S/N threshold of S/N$> 40$, \S \ref{sec:samp}, does not obviously bias our Fnx result in any way, but the chemical abundance patterns overall for Fnx are more uncertain than the other galaxies.

\section{Chemical Evolution Modeling Details}
\label{app:modeling}

As mentioned in the text, we use the two chemical evolution models as a tool to quantify various features in the abundance patterns of these galaxies. There are many uncertainties and degeneracies associated with these models, especially when only fitting median abundance tracks rather than median abundance tracks combined with abundance space density. In this section we provide more detail on some of the parameterization discussed in the text and explore the model parameter space to show how well-constrained the parameters we derive actually are. 

\subsubsection{flexCE Gas Inflow}
\label{flexce:inflow}

One key change we make is that we use a different parameterization of gas inflow than used in the fiducial flexCE model of the Milky Way \citep{Andrews2016}.  We use a delayed tau model for the inflow in each of our models, following the form:  
\begin{displaymath}
\dot{\rm M}_{\rm in} = \bigg(\frac{\rm M_i}{\tau_{\rm i}}\bigg)\bigg(\frac{t}{\tau_{\rm i}}\bigg) e^{-t/\tau_{\rm i}}
\end{displaymath}
where $\dot{\rm M}_{\rm in}$ is the gas mass inflow rate, $\rm M_i$ is the inflow mass scale (with $\rm M_i$ being the total mass that would be accreted as $t \to \infty$), $\tau_{\rm i}$ is the inflow time scale, i.e., the time at which inflow is maximal (although note that this is not necessarily the time at which the star formation rate is maximal), and $t$ is time.  

This form of inflow is motivated by cosmological simulations \citep[e.g.,][]{Simha2014} and is preferred over the fiducial, exponential model of inflow, because the delayed tau model allows for a ramp up of inflow (due to gas accretion earlier in the age of universe), which later cuts off as a Galaxy stops growing through gas accretion and instead would grow through mergers.  Additionally this parametrization of the gas inflow allows us to better reproduce the chemical abundance patterns seen in our sample of dwarf galaxies than when using the fiducial, exponential gas inflow, because it allows for a slower enrichment at earlier times/lower metallicities that helps retain gas for later star formation and enrichment.  In particular, while we don't fit the metallicity distribution functions of these galaxies, or the density of stars in the abundance planes, a slower initial enrichment may be needed to reproduce these quantities once they've been controlled for selection effects.

\subsubsection{flexCE Star Formation Efficiency}
\label{flexce:sfe}

To turn this gas mass into stars, flexCE natively uses a constant star formation efficiency (SFE), such that the star formation rate (SFR) is defined as:  ${\rm SFR} = {\rm SFE} \times {\rm M_{gas}}$.  Here we modify flexCE to include a parametrization of SFE that is time variable in order to be able to simulate a sustained burst of star formation as done in \citet{Nidever2020}, which we employ to fit the chemical abundance profile of the LMC and SMC, as discussed below.  Our formulation is to modify the constant SFE used by flexCE to add an increase in SFE following a Gaussian profile (for simplicity and so that the subsequent SFR change is continuous rather than having jumps or breaks).  Formally, for our burst models, we use a time variable SFE that follows the form:  
\begin{displaymath}
{{\rm SFE} \, (t)} = {\rm SFE} \times \bigg[1+\big({\rm F_b}-1 \big) \, \exp \big({-0.5 \bigg(\frac{(t - \tau_{\rm b})}{\sigma_{\rm b}}\bigg)^2} \, \big) \bigg]
\end{displaymath}
where SFE is the constant base SFE, $\rm F_b$ is the burst strength, i.e., the peak factor of increase of SFE during the burst, $\tau_{\rm b}$ is the time at which the peak increase in SFE occurs, and $\sigma_{\rm b}$ is the scale factor for the duration of the burst.

Of particular note on this parameterization of changing SFE, is that it is not an explicit parameterization of the SFR.  Because the SFR is a function of gas mass and SFE, as the SFE is rises, a galaxy can begin to exhaust its gas reservoir and the SFR may begin to fall, even if SFE continues to rise.  Therefore changing the burst strength, timing, and duration, may have somewhat unintuitive effects on the resulting SFR.  For instance each of these parameters can affect the timing of the SFR burst, because a stronger burst can exhaust gas more quickly and lead to an earlier peak in SFR, as could a longer burst, which may exhaust the model's gas before reaching peak SFE, in addition to simply changing the timing of the peak SFE increase.  Unfortunately, the chemical abundance patterns of galaxies are most sensitive to the SFR of the galaxy, not the underlying SFE, and there are degeneracies among these parameters when it comes to recreating a given SFH, hence we fix two of the three burst parameters when performing our chemical evolution modeling. 

\subsubsection{flexCE Model Sensitivities}

\begin{figure*}
\includegraphics[width=1.0\hsize,angle=0]{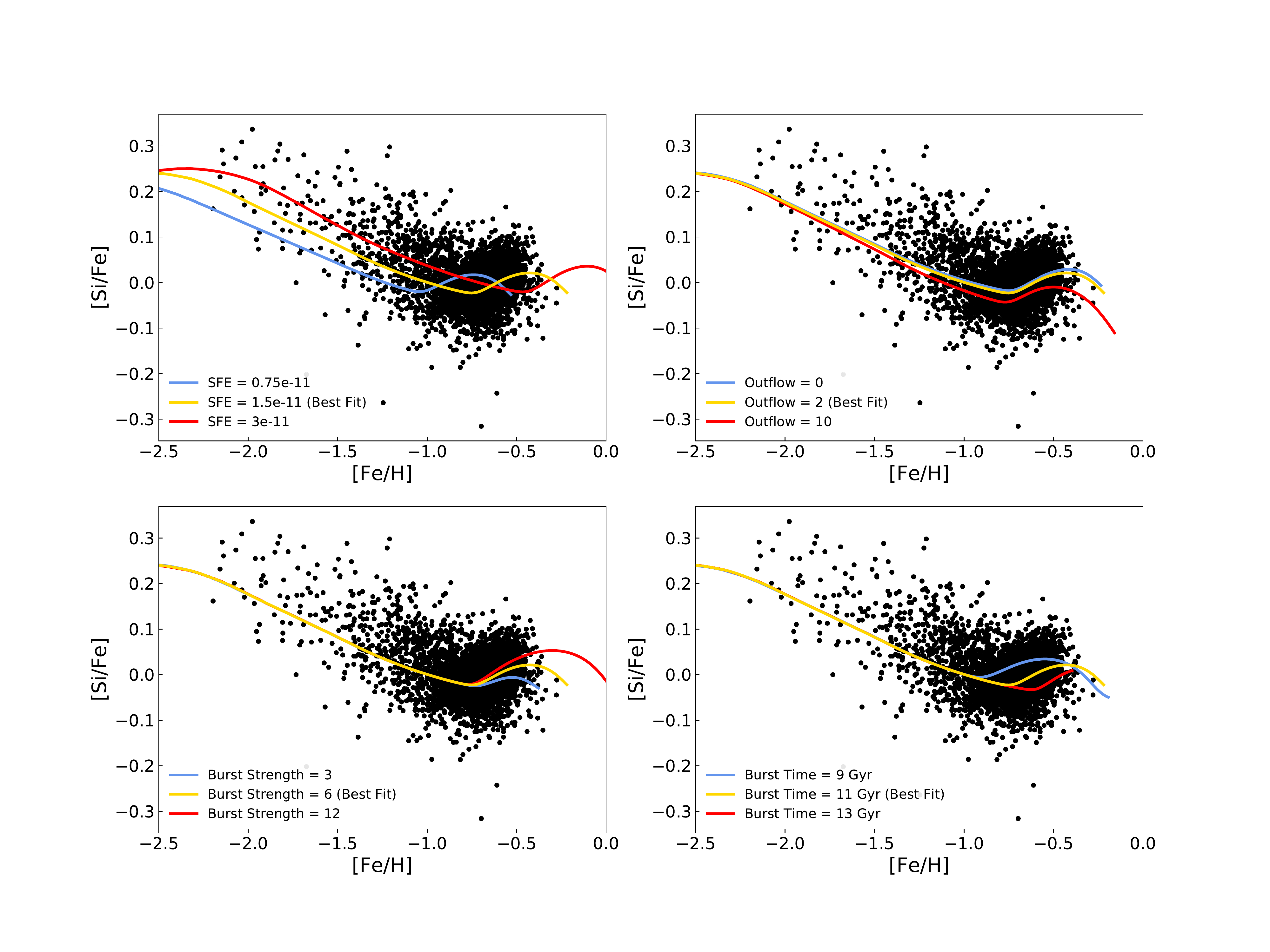}
\caption{The [Si/Fe]-[Fe/H] abundance plane for the LMC stars fit with various flexCE models, where the gold model is the best-fit model presented in \S \ref{sec:chem_results}. Top-left: initial SFE is varied. Top-right: outflow parameter is varied. Bottom-left: burst strength is varied. Bottom-right: time of burst is varied.}
\label{fig:flexce_model}
\end{figure*}

The flexCE modeling has four free parameters that we fit to the median abundance trends. While a detailed $\chi^{2}$ mapping and deriving actual SFH uncertainties is beyond the scope of this work, we show the effects of varying certain model parameters in Figure \ref{fig:flexce_model}. The best-fit LMC model from above is shown in gold, and then models are generated holding all best-fit parameters fixed except for SFE (upper left), outflow strength (upper right), burst strength (lower left), and time of burst (lower right).

The upper-right panel of Figure \ref{fig:flexce_model} shows that while the model prefers a low outflow strength, the model with outflow set to 0 can still reproduce the data fairly well. The strength of the burst, shown in the lower-left panel,  suggests that a stronger burst is not necessarily ruled out by our data, but does make predictions that the youngest LMC stars should have [Fe/H] = 0.0. The lower-right panel shows that the time of burst is reasonably constrained, with an earlier or later burst resulting in a track that does not match the data as well.

\subsubsection{Lian Gas Inflow}
\label{app:lian}

The gas accretion is assumed to decline exponentially, $A(t)=A_{\rm initial}e^{-t/\tau_{\rm acc}}$, where $A_{\rm initial}$ is the initial gas accretion rate and $\tau_{\rm acc}$ is the declining timescale. The major difference between this treatment of the gas inflow and the flexCE treatment is that the Lian model results in stars forming very quickly after the time starts, as much of the gas that will form stars is already present in the galaxy. The flexCE treatment results in a slight delay, as gas must be accreted to push up the star formation rate. This difference in gas inflow could explain some of the differences in SFHs between the two models. Future studies that are able to account for observational biases will be able to use the density of stars in this abundance plane as an additional constraint, perhaps better informing the gas inflow.

\subsubsection{Lian Star Formation Efficiency}

The star formation rate is determined from the gas mass following the form of the Kennicutt-Schmidt star formation law (SFL, \citealt{Kennicutt1998}), assuming a Kroupa IMF \citep{Kroupa2001}. The SFE is thus regulated by the coefficient of the SFL. We assume a constant coefficient ($C_{\rm initial}$) unless a starburst event occurs. The starburst in the Lian model is characterized by an exponential increase in the coefficient of the SFL. In this way, the burst event is described by three parameters, the timescale ($\tau_{\rm burst}$), start time ($t_{\rm start}$) and duration ($\Delta t$) of the SFE increase. After the burst, the coefficient of the SFL is set to decrease exponentially. Since this paper mainly focus on the burst event, for simplicity, we fix this decreasing timescale to be 0.2 Gyr. 

\subsubsection{Lian Model Sensitivities}

\begin{figure*}
\includegraphics[width=1.0\hsize,angle=0]{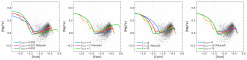}
\caption{The [Mg/Fe]-[Fe/H] abundance plane for the LMC stars fit with various Lian models, where the pink model is the best-fit model presented in \S \ref{sec:chem_results}. Top-left: initial SFE is varied. Top-right: outflow parameter is varied. Bottom-left: burst strength is varied. Bottom-right: time of burst is varied.}
\label{fig:lian_model}
\end{figure*}

Figure \ref{fig:lian_model} shows the results of adjusting various model parameters for the best-fit LMC model. Like the flexCE parameterization discussed above, the Lian model is less sensitive to outflow strength (third panel), but is quite sensitive to the strength of the burst, with stronger or weaker bursts not matching the data (second panel).

\section{Metallicity Distribution Functions}
\label{app:mdfs}

\begin{figure*}
\includegraphics[width=1.0\hsize,angle=0]{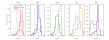}
\caption{Metallicity distribution functions (MDF) of each of the dwarf galaxies along with predicted MDFs from the best-fit chemical evolution models. The flexCE models are denoted by dashed histograms and the Lian models are denoted by dotted histograms.)}
\label{fig:mdf_comp}
\end{figure*}

As discussed in \S \ref{sec:future}, one could use the MDFs to constraint the chemical evolution models. In Figure \ref{fig:mdf_comp} we show the MDFs of the APOGEE data compared to MDFs predicted from the flexCE and Lian best-fit chemical evolution models to illustrate how such comparisons might inform future models. The flexCE-predicted MDFs agree reasonably well with the MC data, whereas the Lian models under-predict the number of metal-poor stars, implying that either the Lian models need to have stronger secondary star formation epochs for the clouds, or that we are biased against observing and/or deriving chemical abundances for the metal-poor stars in the APOGEE sample.

Neither models match well to Sgr or GSE, the two galaxies with the most complicated selection functions. Both models over-predict the number of metal-rich stars in GSE, most likely because they do not truncate star formation at late times. They under-predict the number of metal-rich stars in Sgr, although the Sgr model MDFs agree reasonably well with each other. For some of the APOGEE Sgr core fields, stars were selected as previously known RV members from \citet{Frinchaboy2012}, who specifically targeted M giants in the direction of Sgr. Therefore, it is expected that the APOGEE Sgr sample is somewhat biased against metal-poor stars, perhaps explaining this discrepancy. Both models under-predict the amount of metal-rich stars in Fnx, although the Lian model does predict a relatively larger fraction of metal-rich stars.

\end{appendix}

\end{document}